\documentclass[%
 reprint,
 amsmath,amssymb,
 aps,
 pra,
 showkeys,
]{revtex4-1}
\usepackage[table,x11names]{xcolor}
\usepackage{graphicx}
\usepackage{dcolumn}
\usepackage{bm}
\usepackage{hyperref}

\usepackage{cleveref}
\usepackage{float}

\begin{document}


\title{A distance constrained synaptic plasticity model of \textit{C. elegans} neuronal network}



\author{Rahul Badhwar$^1$}
\author{Ganesh Bagler$^{1}$$^,$$^2$$^,$$^3$}%
\email{bagler@iiitd.ac.in}
\affiliation{$^1$Centre for Biologically Inspired System Science, Indian Institute of Technology Jodhpur, Jodhpur, Rajasthan, India}
\affiliation{$^2$Dhirubhai Ambani Institute of Information and Communication Technology, Gandhinagar, Gujarat, India}
\affiliation{$^3$Center for Computational Biology, Indraprastha Institute of Information Technology Delhi (IIIT-Delhi), New Delhi, India}


\begin{abstract}
Brain research has been driven by enquiry for principles of brain structure organization and its control mechanisms. The neuronal wiring map of \emph{C. elegans}, the only complete connectome available till date, presents an incredible opportunity to learn basic governing principles that drive structure and function of its neuronal architecture. Despite its apparently simple nervous system, \emph{C. elegans} is known to possess complex functions. The nervous system forms an important underlying framework which specifies phenotypic features associated to sensation, movement, conditioning and memory. In this study, with the help of graph theoretical models, we investigated the \emph{C. elegans} neuronal network to identify network features that are critical for its control. The `driver neurons' are associated with important biological functions such as reproduction, signalling processes and anatomical structural development. We created 1D and 2D network models of \emph{C. elegans} neuronal system to probe the role of features that confer controllability and small world nature. The simple 1D ring model is critically poised for the number of feed forward motifs, neuronal clustering and characteristic path-length in response to synaptic rewiring, indicating optimal rewiring. Using empirically observed distance constraint in the neuronal network as a guiding principle, we created a distance constrained synaptic plasticity model that simultaneously explains small world nature, saturation of feed forward motifs as well as observed number of driver neurons. The distance constrained model suggests optimum long distance synaptic connections as a key feature specifying control of the network.
\end{abstract}

\maketitle


\section{Introduction}

The quest for understanding broad structural organization, functional building blocks and mechanisms of control of nervous systems has been central to neuroscience~\cite{kandel2000principles}. Vast knowledge of cellular and molecular mechanisms garnered through reductionist studies over decades, while enriching our understanding of brain mechanisms, have highlighted the need for holistic perspective of neural architecture~\cite{sporns2011networks}. This urge to delve into systems properties has propelled efforts into connectome projects that attempt to map and model neural wirings to the finest detail possible~\cite{White1986, Chiang2011, Sporns2013, Zingg2014}. \textit{C. elegans} connectome is the only complete neuronal wiring diagram available till date~\cite{White1986, Chen2006, Towlson2013}. Along with the rich understanding available on the biology of this model organism~\cite{altun2002wormatlas, Howe2016}, its connectome presents an opportunity to learn basic governing principles that drive structure and function of neuronal architecture.

Despite its apparently simple nervous system, \textit{C. elegans} is known to possess complex functions associated to sensation, movement, conditioning and memory~\cite{Chatterjee2008, Ardiel2010}. This multi-cellular nematode has been extensively investigated to understand neural mechanisms involved in response to chemicals, temperature, mechanical stimulation as well as mating and egg laying behaviors~\cite{Hobert2003, Chatterjee2008}. These biological functions have neuronal basis and are a reflection of emergent properties of signal dynamics over the network. Its nervous system has evolved to confer evolutionary benefits under constant tinkering and is known to undergo synaptic rewiring during the course of its life~\cite{Kandel2014}. Beyond the broad evolutionary architecture, synaptic plasticity offers additional adaptive advantage to respond to the environment and perhaps to achieve better functional efficiency. The key role of distance constraint in shaping the architecture of complex networks has been well studied and highlighted~\cite{Amaral2000, Barthelemy2011, Avena-Koenigsberger2015}.

\begin{figure}[!t]
\includegraphics[scale=0.52]{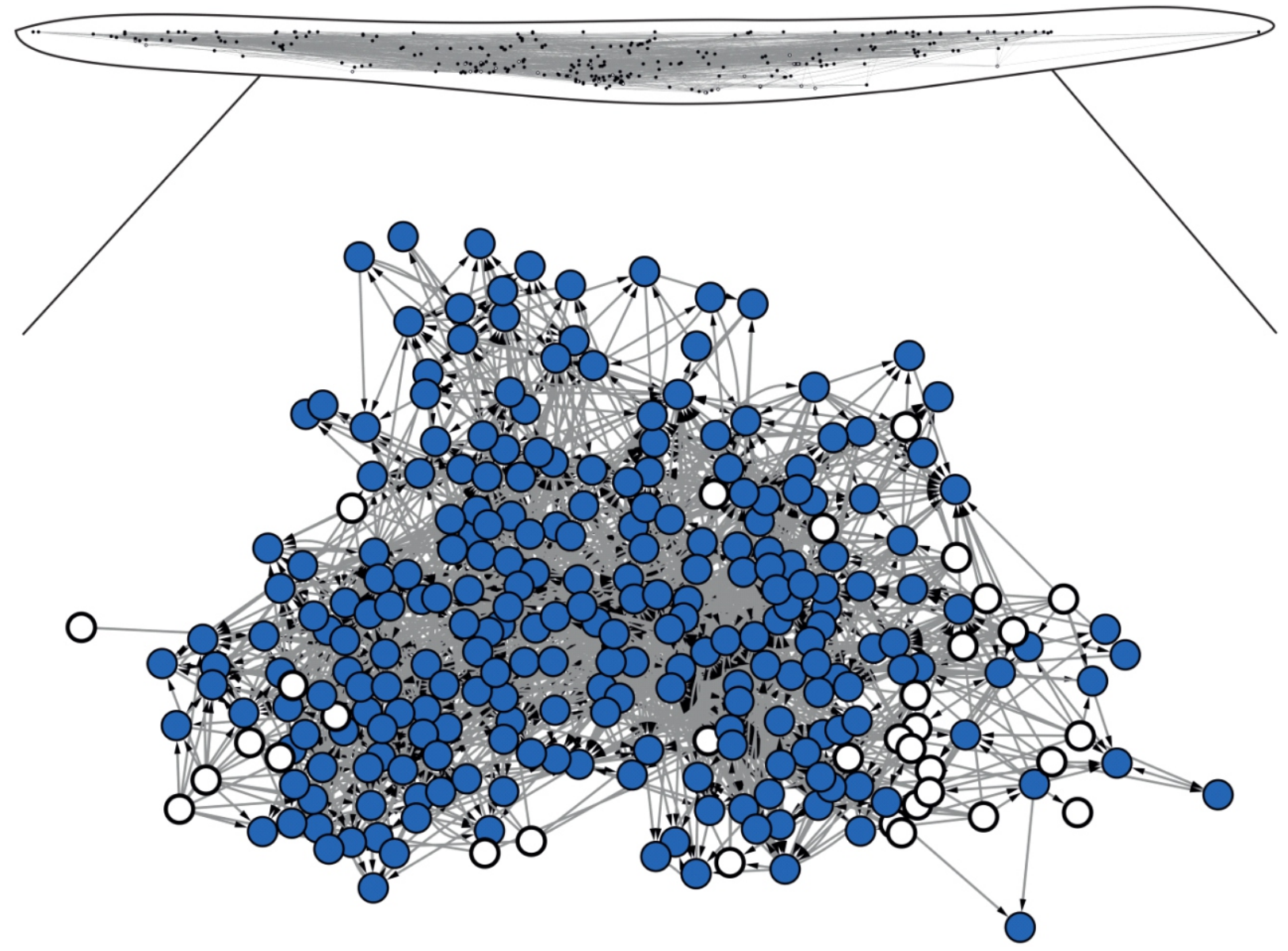}
\caption{Network structure of \textit{C. elegans} nervous system. Functionally relevant driver neurons (34 nodes highlighted in white) were identified with maximum matching criterion. Beyond explaining the small world nature, saturation of feed forward motifs and observed number driver neurons, the distance constrained synaptic plasticity model accurately identifies specific driver neurons.}
\label{fig:cenn_network}
\end{figure}

The \textit{C. elegans} neuronal system could be modelled as a network and studied for structural properties of its neuronal architecture as well as for network dynamics~(\figurename~\ref{fig:cenn_network}). Graph theoretical studies provide important insights into evolutionary mechanisms of this system and enable biological inference. The \textit{C. elegans} neuronal network (CeNN) has been mapped to a high resolution with details of its neurons, their locations and synaptic connectivity~\cite{Choe2004}. The network, comprising of 277 neurons that are interlinked with 2105 synapses, has been studied for its broad structural features as well as towards identification of motifs that potentially contribute to the dynamics over the network. Using graph theoretical measures, CeNN has been observed to have a small world architecture with small path length and high clustering~\cite{Watts1998}. One of the mechanisms by which interconnected systems acquire small world nature is by having modules densely connected with short-range connections, which are further interlinked through long-range connections~\cite{Watts1998}. It has been proposed that such connectivity pattern may emerge due to processes that leave the network critically poised between absolute order and extreme randomness. The small world nature may render this neural network (as well as other neuronal systems) efficient for information dynamics. Such a topology is known to offer evolutionary advantage by optimized wiring in neuronal systems~\cite{Chen2006, Ahn2006}.

Networked systems are known to be built with recurring circuit modules that are central to their function~\cite{Alon2007}. When probing for network sub-structures that could form the building block of the CeNN, Milo \textit{et al.}~identified feed forward motifs (FFMs) to be significantly over-represented~\cite{Milo2002}. FFMs have been suggested to be of functional relevance to biological systems such as transcriptional regulatory networks~\cite{Mangan2003, Balaji2007}. One of the possible utilities ascribed to these structural building blocks is control of signal regulation in response to persistent input. How exactly such building blocks may offer functional advantage to neuronal networks and whether these entities have evolved to optimize the building blocks is not clearly understood yet.

Control systems approach to complex networks provides a better perspective of dynamics over the network and ability to steer its `state'~\cite{Liu2011}. Neuronal architecture of CeNN forms an important underlying framework which specifies phenotypic features of \textit{C.\ elegans}. Important behavioral traits as well as cognitive processes (such as movement, sensation, egg laying, mechanosensation, chemosensation and memory) are known to have neuronal basis. A network is said to be controllable if it can be reached to a desired state from any initial state by providing inputs to certain nodes~\cite{Lin1974, Liu2011}. The set of nodes that facilitate such a control are named driver nodes~\cite{Liu2011}.

By studying genotypic and phenotypic aspects of CeNN, in our earlier study we have shown that `driver neurons' are associated with important biological functions such as reproduction, signalling processes and anatomical structural development~\cite{Badhwar2015}. Going by their relevance to structural controllability, driver neurons are expected to be important in dictating the state of the network. In \textit{C. elegans} driver neurons are primarily of short span and linked to motor activities~\cite{Badhwar2015}. Interestingly, randomized controls have no driver neurons as compared to CeNN which presents a sizeable number of driver neurons that are crucial for its control. While earlier studies have shown that connectivity of neurons in CeNN partially explains the observed number of driver neurons~\cite{Liu2011}, no model has so far been developed that accounts for its small world architecture, over-representation of FFMs as well as controllability.

In this study, we create one-dimensional (1D) and two-dimensional (2D) network models of \textit{C. elegans} neuronal system to investigate the role of FFMs as building blocks in conferring controllability and small world nature. With the help of a simple 1D ring model we show such a network is critically poised for the number of FFMs, neuronal clustering and characteristic path-length in response to synaptic rewiring, indicating optimal rewiring. We found that synaptic connections between neurons are characterized with a strong distance constraint in CeNN. Using this as a guiding principle, we created a distance constrained synaptic plasticity model that simultaneously explains small world nature, FFM saturation and controllability of the network. This model could account for the number of driver neurons observed in CeNN.
Moreover, the nodes that act as driver neurons in this model match with those obtained from empirical network with high accuracy. Thus the model highlights realistic process of distance constrained synaptic plasticity as a plausible basis of nature of functional sub-structures and controllability observed in CeNN.

\section{Materials and methods}

\subsection{\textit{C. elegans} neuronal network}
The nervous system of \emph{C. elegans} consists of 277 neurons (barring the pharyngeal neurons) which are interconnected via electrical and chemical synapses~\cite{White1986, Choe2004}. We constructed CeNN, a graph theoretic model of \emph{C. elegans} neuronal network, comprising of 277 somatic neurons and 2105 synaptic connections. Multiple synaptic connections between two neurons were merged to yield a simple directed unweighted graph in which neurons represent nodes and synaptic connections are links. A typical neuron in CeNN on an average had $7.59$ synaptic connections.

\subsection{Topological properties of CeNN}
We calculated following graph theoretical properties of the network embodying clustering, compactness, structural motifs and controllability of the network.

\subsubsection{Clustering coefficient} Clustering coefficient of a node $C_{i}$ is defined as ratio of number of triangles (triangle refers to a three node clique) made by a node with its neighbours to the maximum number of triangles that can be formed by them~\cite{Watts1998}. For a graph $G=(V,E)$ the clustering coefficient of a node $i$ is defined as follows:

\begin{equation*}
C_{i} = \dfrac{\mid \lbrace e_{jk}:v_{j}, v_{k} \in N_{i}, e_{jk} \in E \rbrace \mid}{k_{i}(k_{i}-1)}
\end{equation*}

Here, $N_i$ refers to the neighbourhood of node $i$ and $k_i$ represents its connectivity (degree).

The average clustering coefficient ($\overline{C}$) was calculated by averaging clustering coefficients of all $n$ nodes: \mbox{$\overline{C} = \frac{1}{n}\sum_{i=1}^{n}C_{i}$.}

\subsubsection{Characteristic path-length}
Characteristic path-length~($L$) enumerates compactness, reflecting ease of information transfer, of the network. It is defined as the average of shortest path-lengths among all pairs of nodes in the network.

\begin{equation*}
L = \dfrac{1}{n(n-1)} \cdot \sum_{i\neq j} d(v_{i}, v_{j})
\end{equation*}

\subsubsection{Feed forward motifs}
Network motifs are defined as patterns of interconnections occurring in complex networks at numbers that are significantly higher than those in randomized networks~\cite{Milo2002}. In a three node digraph $13$ different types of three node motifs can exist. Angular motifs are linear three node sub-structures, and triangular motifs comprise of three nodes inter-connected with either unidirectional or bidirectional edges. For our studies, we computed number of feed forward motifs, $n_{FFM}$, (among  unidirectional triangular motifs) that are prevalent in many real world networks including CeNN~\cite{Milo2002, Alon2007}. Please see Section~S1 (\figurename~S1 and \figurename~S2) of Supplementary Material for more details. We used the algorithm employed by Milo \emph{et al.} for identification and enumeration of frequency of occurrence motifs~\cite{Milo2002}. The $Zscore$, indicating significance of observed number of FFMs in CeNN, was calculated by comparing it with random controls: $Zscore = \frac{n_{FFM} (CeNN)- \overline{n_{FFM}(ER)}}{ \sigma_{ER}}$.

\subsubsection{Number of driver neurons}
From control systems perspective, driver nodes in a network are those nodes which when controlled by an external input can provide full control over the state of the network~\cite{Liu2011}. Analogously, we term driver nodes in CeNN as  driver neurons. Due to their role in control of network, driver neurons are of functional relevance to the neuronal network~\cite{Badhwar2015}. We computed minimum number of driver neurons $(n_{D})$ using maximum matching criterion~\cite{Liu2011}. A node is said to be matching if any matching edge is pointing towards it and is unmatched if no matching edge is directed towards it. We implemented maximum matching algorithm proposed by Pothen \emph{et al.} to find unique unmatched nodes by augment matching~\cite{Pothen1990}.

\subsection{Random controls of CeNN}
We constructed two random controls of CeNN viz.\ Erd\"{o}s-R\'{e}nyi random control (ER) and degree distribution conserved control (DD)~\cite{Erdos1959, Maslov2002}. In ER control, number of nodes and edges were kept the same as that of CeNN but the connectivity was random. In DD control, the in-degree and out-degree of each node was also preserved in addition to number of nodes and edges.

\begin{figure}[!t]
\centering
\includegraphics[scale=0.2]{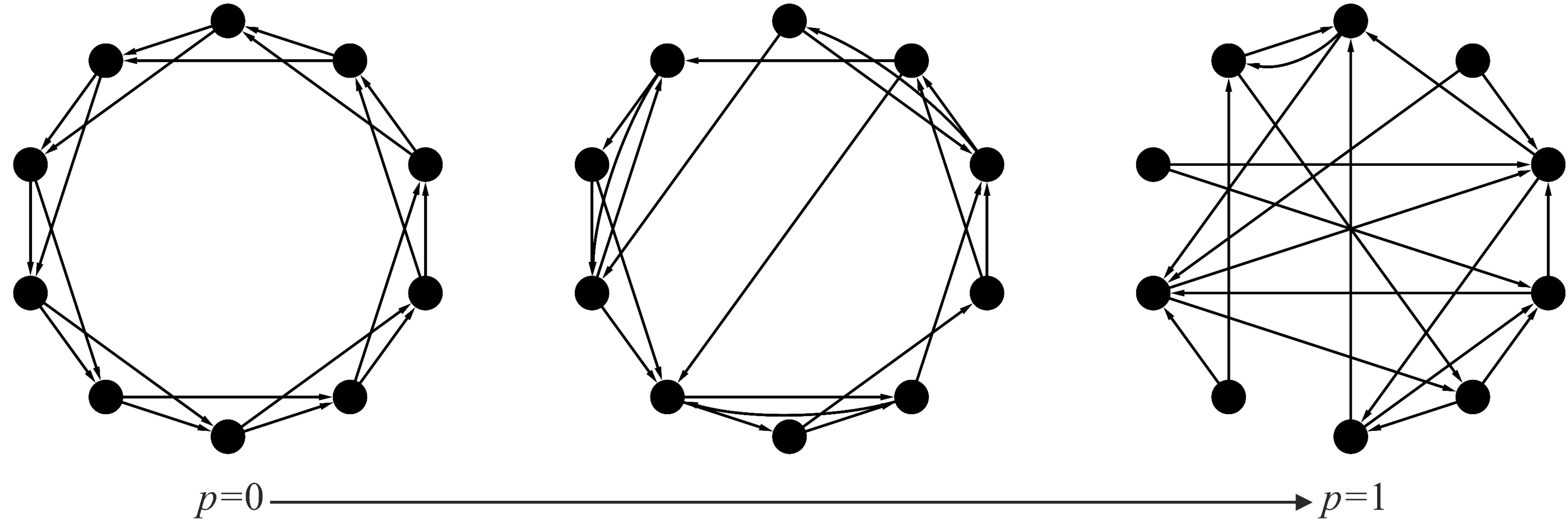}
\caption{The 1D ring model, with neurons linked for maximizing number of feed forward motifs, was rewired with increasing probability of synaptic rewiring. Starting with an asymptotic model (with 277 nodes and 8 out-going edges) saturated with FFMs, synaptic rewiring was emulated with probability $p$. The model exhibits a spectrum of topological variations between extreme regularity and randomness. The figure shows an illustration for 10 nodes and 2 outgoing edges. See \figurename~S3 of Supplementary Material for another illustration.}
\label{fig:regular_graph}
\end{figure}

\subsection{1D ring model of CeNN}
We constructed a ring graph model of CeNN so as to maximise the number of FFMs while preserving the number of neurons $(n)$ as well as average neuronal connectivity $(k)$ of CeNN~(\figurename~\ref{fig:regular_graph}).
While the core idea and strategy implemented in this model is analogous to that of Watts and Strogatz's~\cite{Watts1998}, it is extended to represent directed edges (synapses), and hence naturally accommodates network motifs and controllability analysis.
Starting with $n(=277)$ nodes arranged in a circular manner, every node was connected (in anti-clockwise sense) to its next nearest neighbour with a directed edge. The procedure was repeated to connect every node with its nearest neighbour and the next nearest neighbour until the out-degree of every node matched with that of average out-degree of CeNN $(k=7.59 \approx 8)$. This strategy maximises the number of FFMs to ${nk(k-1)}/{2}$ in the regular graph model of CeNN and represents an asymptotic version saturated with FFMs.

To mimic random synaptic plasticity in this simple 1D model, we rewired every edge in this network with a certain `probability of rewiring ($p$)'. Every out-going edge connecting a node to its nearest neighbour was chosen and rewired randomly with probability $p$ by ensuring that there were no duplicate edges or self-edges and that the network is always connected. In the second lap, the process was repeated for the edges made with next-nearest neighbours and so on. All edges are thus exhaustively considered for rewiring in $k$ laps. For every probability of rewiring $1000$ instances of graphs were created for a range of $p=10^{-4}$ to $p = 1$. Topological properties ($\overline{C}$, $L$, $n_{FFM}$ and $n_D$) were computed for every instance of graph thus generated.

\subsection{Distance constrained models of CeNN}
We created 2D distance constrained models that, similar to distance constraint observed in CeNN, follow a restraint on synaptic connectivity based on distance between two neurons. These 2D models are based on positional data of \emph{C. elegans} neurons, that have been mapped to a high resolution~\cite{Choe2004}. In these models, the probability $P(d)$ that two neurons at a distance $d$ are connected with a synapse approximately follows a power law pattern observed from  empirical data:
\begin{equation*}
P \propto d^{- \alpha}
\end{equation*}

The distance constraint is modulated by the exponent $0 \le \alpha \le \infty$. Here, the distance between neurons $i$ and $j$, $d(i,j)$, was calculated as the Euclidean distance: \mbox{$d(i,j) = \sqrt{(x_{i} - x_{j})^{2} + (y_{i} - y_{j})^{2}}$.} The power law nature of neuronal connectivity was established following the recipe suggested by Clauset \textit{et al.}~\cite{Shalizi2009}.

We created two models of CeNN based on the distance constraint: Distance constrained random (DCR) and Distance constrained synaptic plasticity (DCP).

\subsubsection{Distance constrained random (DCR) model}
The underlying framework for DCR model is that of ER control. Starting with ER (random) control, we rewired every edge to impose distance constraint for specific exponent $\alpha$. Statistics of topological parameters were computed over 100 instances. Response of DCR model was observed by varying the value of exponent between $0 \leq \alpha \leq 3$.

\subsubsection{Distance constrained synaptic plasticity (DCP) model}
In contrast to DCR model, the underlying framework for DCP model is that of DD control which preserves the synaptic connectivity of each neuron. Starting with DD control, every edge was rewired to impose distance constraint for specific exponent $\alpha$ ($0 \leq \alpha \leq 3$) and statistics of topological parameters were computed over 100 instances.

\subsection{Cartesian graph model of CeNN}
The deterministic Cartesian graph model of CeNN was created by ensuring that every neuron is connected to its spatially nearest neurons. Beginning with $(n=)$ 277 neurons placed at cartesian coordinates matching their observed position in the nervous system of \emph{C. elegans}~\cite{Choe2004}, every neuron was connected to $(k=)$ 8 spatially nearest neurons. This model reflects preferential deterministic connections made by a neuron based on its distance from another neuron.

\subsection{Identification of specific driver neurons}
Using maximum matching algorithm~\cite{Pothen1990}, the set of specific driver nodes was identified from both the empirical neuronal network as well as its computational models (DCR and DCP). The latter were compared against the former to assess the performance of models in achieving real-like topology and control structure. The success of DCR, DCP models in accurate identification of driver neurons was measured with the help of F1 score. Using the driver neurons set identified from the CeNN (34) as the basis (Details of driver neurons is provided in Table~S3 of Supplementary Material), we identified true positives~$(TP)$ and true negatives~$(TN)$ (neurons that are correctly classified) as well as false positives~$(FP)$ and false negatives $(FN)$~(neurons that were incorrectly marked as driver neurons, and neurons that were incorrectly marked as non-driver neurons, respectively) for DCR and DCP models across 100 instances. The $F1$ score, which is used for computing the quality of binary classification is defined as,~$F1=\frac{2~TP}{2~{TP} + {FP} + {FN}} $.

\section{Results}

\subsection{Topological properties of CeNN}
Topological features of network provide insights into its structure and function~\cite{Albert2002, dorogovtsev2010lectures}. Consistent with previous reports, we observed that \textit{C. elegans} neuronal network is a small world network by virtue of high clustering coefficient~($\overline{C} = 0.172$) and comparable characteristic path length~($L = 4.018$), with respect to its randomized counterpart ($\overline{C}_{ER} = 0.028$ and $L_{ER} = 2.97$)~\mbox{(Table~\ref{Tab:diff})}~\cite{Watts1998}. Beyond these global topological features, CeNN is known to be over represented with feed forward motifs~\cite{Milo2002} that are functionally associated with mechanisms of memory~\cite{Mozzachiodi2010}. We observed that, FFMs were significantly overrepresented in CeNN $(Zscore = 151.12)$ as compared to those in corresponding random graphs.

\begin{table}[!t]
\centering
\renewcommand{\arraystretch}{1.2}
\caption{Topological properties of CeNN and its controls.\label{Tab:diff}}
\begin{tabular}{|c||c||c||c|}\hline
 & CeNN & ER & DD\\ \hline
$\overline{C}$ & 0.172 & 0.028 $\pm$ 0.001 & 0.067 $\pm$ 0.003\\ \hline
$L$ & 4.018 & 2.97 $\pm$ 0.01 & 2.981 $\pm$ 0.018\\ \hline
$n_{D}$ & 34 & 0.28 $\pm$ 0.514 & 22.38 $\pm$ 1.153\\ \hline
$n_{FFM}$ & 3776 & 438.3 $\pm$ 22.1 & 1699.6 $\pm$ 57.5\\ \hline
\end{tabular}{}
\end{table}

From control systems perspective CeNN can be controlled through a small set of driver neurons (34) to any desired state in finite time~\cite{Liu2011}. The number of driver neurons in CeNN is significantly higher in comparison to its random counterpart. Driver neurons in CeNN are genotypically and phenotypically associated with biological functions such as reproduction and maintenance of cellular processes~\cite{Badhwar2015}. This alludes to the fact that driver neurons serve a critical role in the neuronal architecture of \textit{C. elegans} and the number of driver neurons therefore has functional bearing on its control.

Table~\ref{Tab:diff} depicts topological features of CeNN that are potentially critical for specifying its function. Other than the small world nature, evident from high clustering among neurons, the CeNN is characterized with significantly higher number of driver nodes as well as number of feed forward motifs. While connectivity (degree) of neurons~(DD) partially explains the increase in FFMs as well as that in $n_{D}$, at the same time it cannot account for observed clustering. No comprehensive model that can explain all of these functionally relevant features is hitherto known.

\subsection{1D ring model of CeNN}

\begin{figure}[!t]
\centering
\includegraphics[scale=0.2]{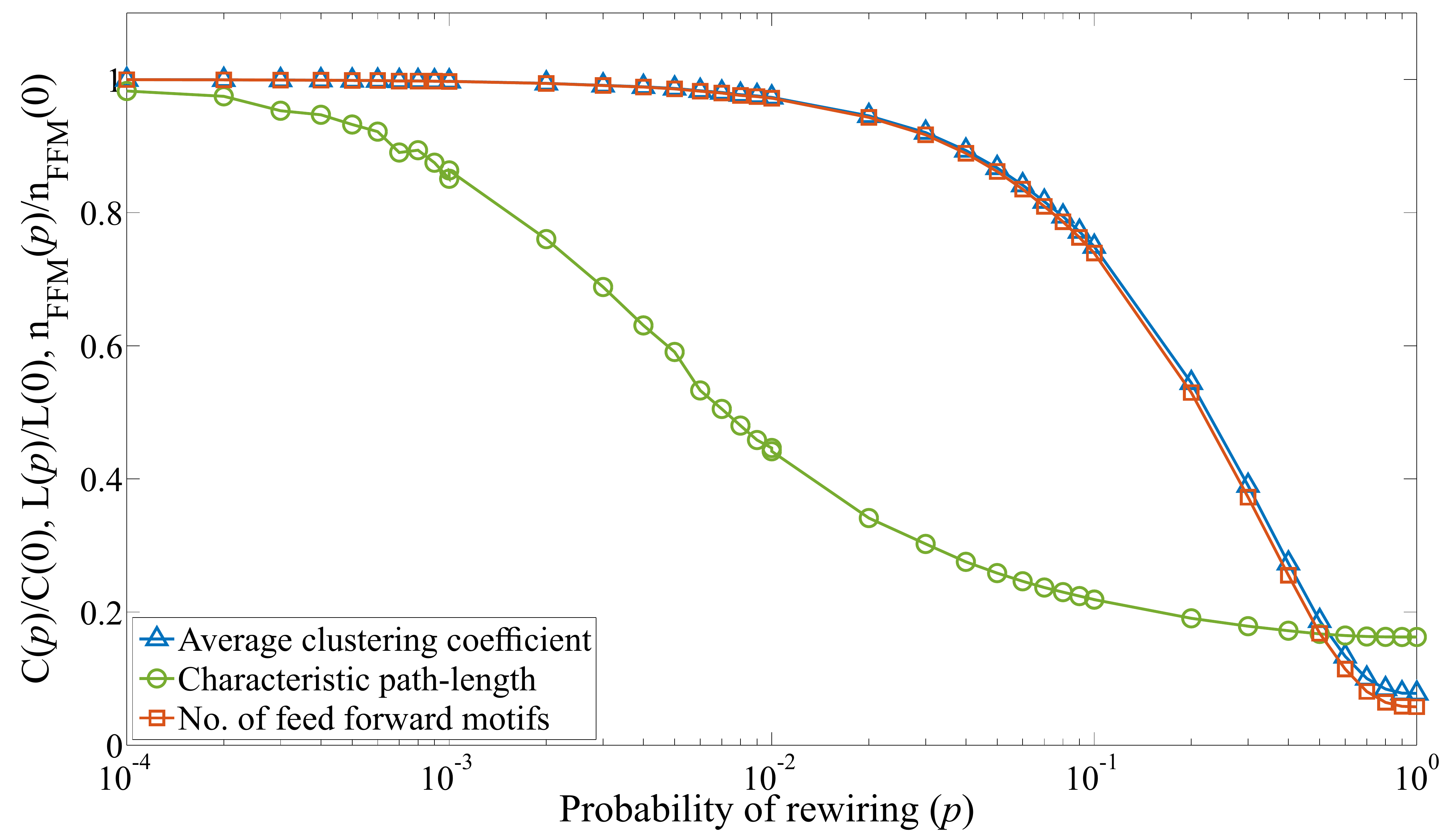}
\caption{Response of 1D ring model with changing probability of synaptic rewiring was measured in terms of average clustering coefficient~($\overline{C}$), characteristic path-length~($L$),  number of FFMs~($n_{FFM}$) and number of driver nodes~($n_D$). For intermediate values of $p$, the model exhibits small world phenomenon as well as FFM saturation, but can not account for controllability ($n_D=0$~$\forall$ $p$). All parameters were normalized with respect to the initial ring graph ($p=0$). Error bars represent standard deviation over 100 instances. Please see Section~S3 of Supplementary Material (\figurename~S4, \figurename~S5 and \figurename~S6) for non-normalized data.}
\label{fig:cenn_p_rewire}
\end{figure}

To investigate for possible mechanisms that could have lead to the observed saturation of FFMs, we created a 1D ring model of CeNN. A directed regular ring graph with $n$ neurons and $k$ average synapses can have a maximum of $nk(k-1)/2$ FFMs as shown in (\figurename~\ref{fig:regular_graph}). Starting from such a regular ring graph maximally saturated with $7756$ FFMs, we simulated random synaptic rewiring to observe its effect on topological features. In addition to FFM saturation, the regular graph had very high average clustering coefficient $(\overline{C}_{reg} = 0.35)$ as well as characteristic path-length $(L_{reg} = 17.69)$. From an analogous undirected Watts and Strogatz model it was anticipated that with increase in synaptic rewiring the clustering as well as path-length would decrease to approach that of random graph asymptotically~\cite{Watts1998}. This simulation of synaptic rewiring was also expected to provide insights into its impact on number of FFMs and driver neurons. As shown in the \figurename~\ref{fig:cenn_p_rewire}, with increasing probability of synaptic rewiring the number of FFMs is unaffected up to $p\approx 0.01$ before falling sharply. While this result points at a critical threshold for number of FFMs in response to probability of synaptic rewiring, no driver neurons were presented by the model across the simulation ($n_{D} \approx 0$ ~ $\forall$ ~ $0 \leq p \leq 1$). For 1D ring graph, these results highlight a critical threshold of rewiring for which the network has optimum saturation of FFMs. This implies that to reflect small world nature and saturation of FFMs, the 1D representation of CeNN would need to have an optimum extent of rewiring.
Such a simple model can only provide topological insights devoid of biological basis and clearly can not justify observed controllability. Search for a more realistic model prompted us to look for biological constraints that may dictate synaptic rewiring as well as to build a 2D model that could possibly reveal mechanisms that render observed controllability in CeNN.

\subsection{CeNN follows distance constrained synaptic connectivity pattern}
\begin{figure}[!t]
\centering
\includegraphics[scale=0.2]{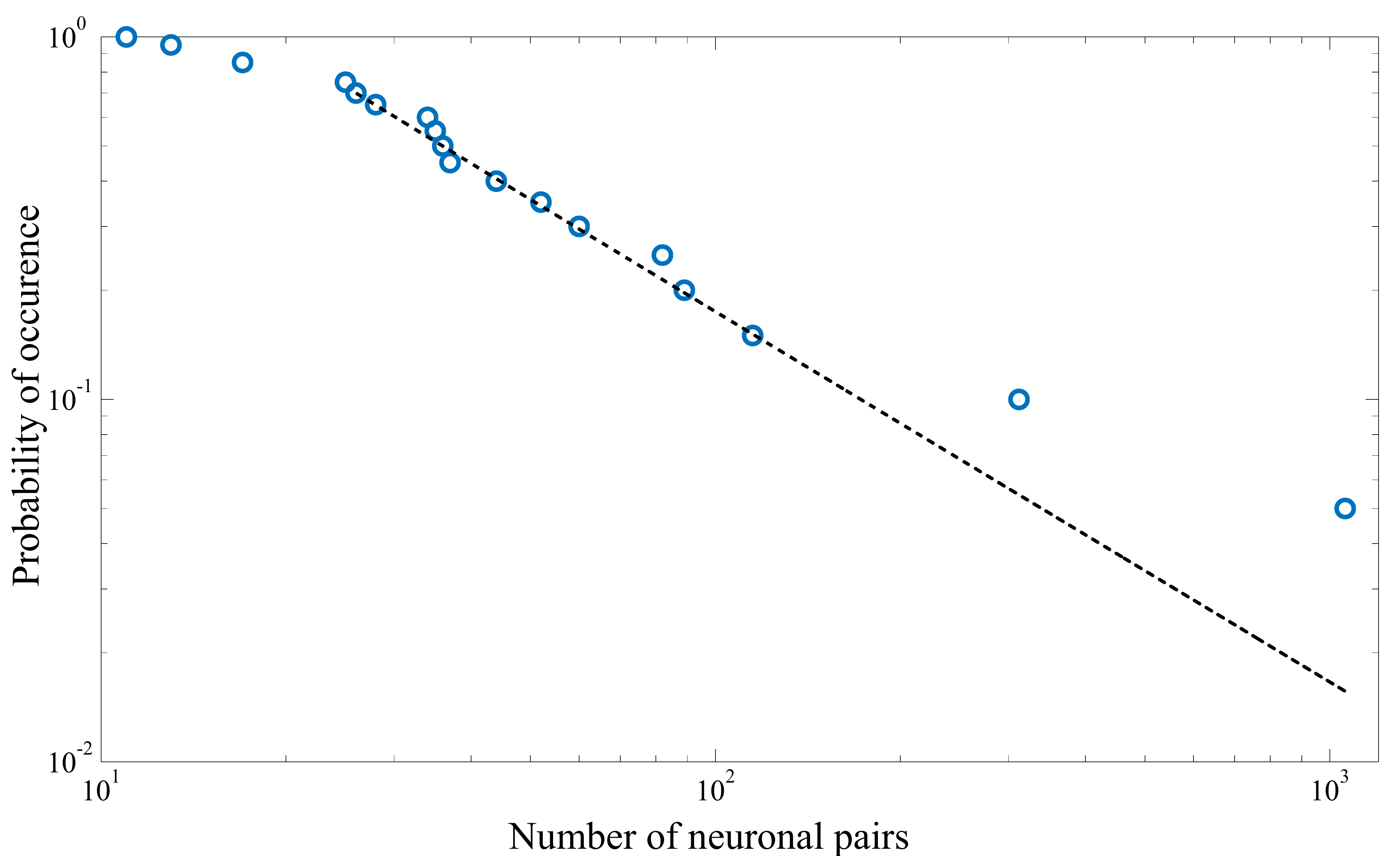}
\caption {Empirical distance constrained synaptic connectivity pattern observed in \textit{C. elegans} neuronal wiring. The number of synapses that connect neurons at distance $d$ follows a power law pattern with an exponent of $\alpha=2.02$ ($p$-value=$0.92$)~\cite{Shalizi2009}.}
\label{fig:dist_pair_log}
\end{figure}

We measured the connectivity pattern in CeNN by enumerating number of neuron pairs that are synaptically connected and cartesian distance between them. We observed that the synaptic connections were constrained by distance as evident from the power law observed from neuronal connectivity data~(\figurename~\ref{fig:dist_pair_log}). The probability of two neurons being connected scales as a power law~($P(d) \propto d^{- \alpha}$), with presence of a few exceptional long distance connections with an exponent $\alpha=2.02$ ($p$-value=$0.92$). The power law nature of data was established following the strategy prescribed by Clauset \emph{et al.}~\cite{Shalizi2009}.

\begin{figure*}[!t]
\centering
\includegraphics[scale=0.42]{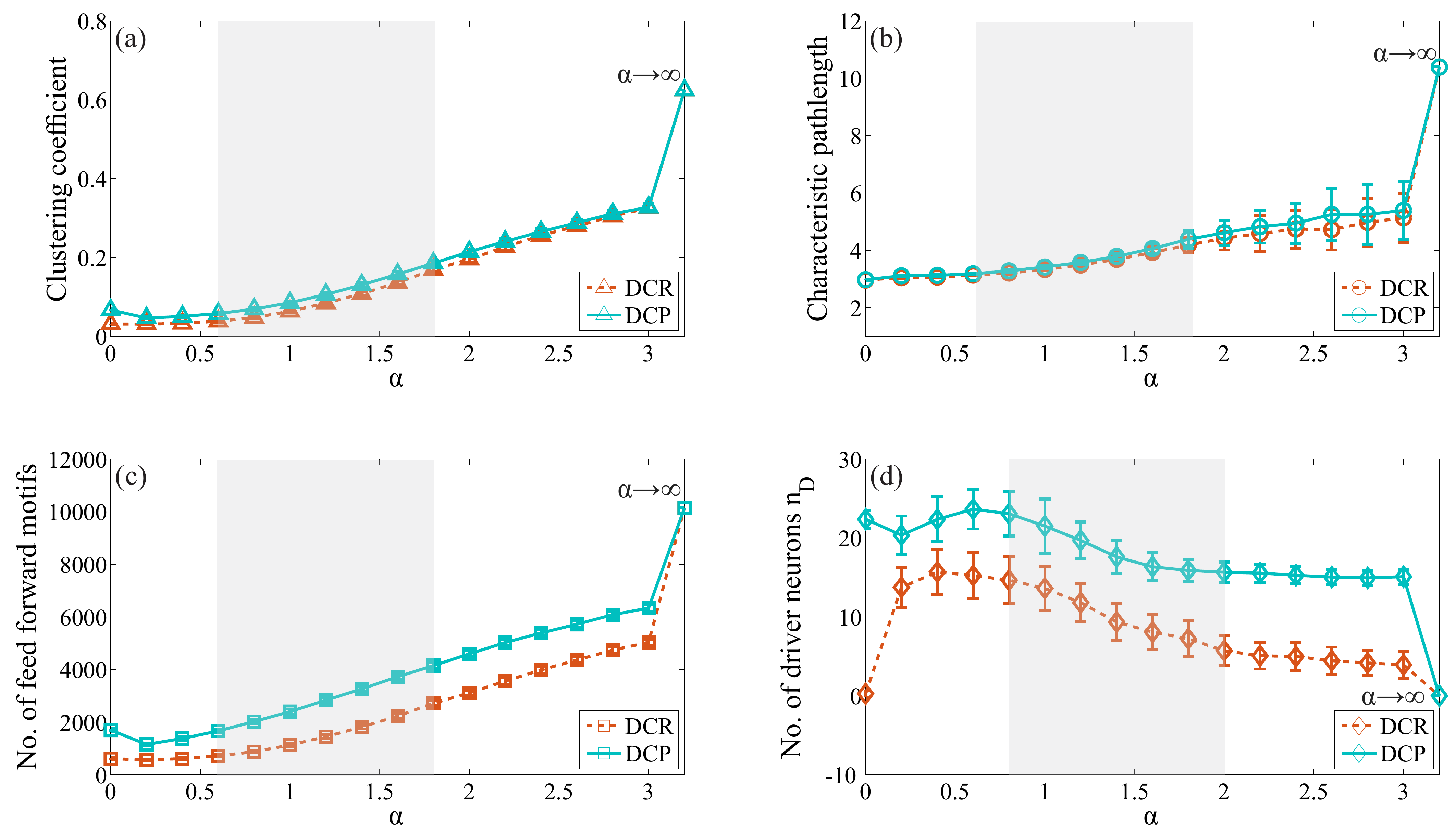}
\caption{Response of distance constrained models of CeNN (DCR and DCP) with increasing constraint ($\alpha$) measured in terms of (a) average clustering coefficient~($\overline{C}$), (b) characteristic path-length~($L$),  (c) number of FFMs~($n_{FFM}$), and (d) number of driver nodes~($n_D$). The lower the $\alpha$ more heterogeneous are the synaptic lengths (larger proportion of long range synapses). For $\alpha = 0$, DCR and DCP models converge to ER and DD controls, respectively. For asymptotic limits of $\alpha \rightarrow \infty$ both the models converge to the Cartesian model, a regular model with saturation of FFMs coupled with high clustering but devoid of driver nodes. While the small world nature (reflected in high clustering and low path-length) and FFM saturation is realized by both DCR and DCP models, DCP model stands out in reproducing all key features of CeNN for $0.6 \leq \alpha \leq 1.8$ (highlighted with gray background). Please see Figure~S7, Table~S1 and Table~S2 of Supplementary Material for details about nature of degree distributions and associated data.}
\label{fig:dcr_dcp}
\end{figure*}

To incorporate this empirical distance constrained connectivity pattern we created more realistic 2D models: (1) Distance constrained random (DCR) model that adds distance constraint starting from ER control, and (2) Distance constrained plasticity (DCP) model that overlays the distance constraint starting from the DD control. Along with the ER and DD random controls these models allow us to segregate the contribution of degree (connectivity) vis-\`a-vis distance constrained synaptic wiring towards conferring observed topological features upon CeNN.

\subsection{Distance constrained random model}
The distance \mbox{constrained} random (DCR) model is a 2D model in which the number of neurons, number of synapses and neuronal locations were preserved. Starting from initial random connectivity (ER) every synapse was probabilistically rewired to follow distance constraint with a certain $\alpha$~(\figurename~\ref{fig:dcr_dcp}). The lower asymptotic limit of this model converges to ER model for $\alpha = 0$. With increasing $\alpha$ the probability of long distance synaptic connections decreases. For extremely large values of $\alpha$ this model converges to the Cartesian model in which every neuron is deterministically connected to its spatially nearest neighbours. We varied the value of $\alpha$ between $0$ and $3$ to assess its impact on the topology of neural network. We found that the average clustering coefficient, characteristic path-length and number of FFMs monotonously increase with increasing $\alpha$. For these parameters the DCR model was closest to actual neuronal network of \textit{C. elegans} for $\alpha = 0.6$. While this model took us closer to CeNN, it did not reflect controllability measured in terms of $n_{D}$. The driver neurons vanish for asymptotic limits of $\alpha$ with maximum $n_{D} = 15.7$ for $\alpha = 0.4$. The fact that DD control, in which number of synapses of every neuron is preserved, matches with CeNN better in controllability~(Table~\ref{Tab:diff}) prompted us to create a more refined `distance constrained synaptic plasticity model'.

\subsection{Distance constrained synaptic plasticity model}
The distance constrained synaptic plasticity (DCP) model preserves the number of synapses of every neuron over and above the number of neurons and their locations. While following the distance constraint, this model mimics synaptic rewiring that is known to take place in CeNN~\cite{Shen2003, Kandel2014}. We observed the response of topological features for varying extent of distance constraint $(0 \leq \alpha \leq 3)$~(\figurename~\ref{fig:dcr_dcp}). In addition to the clustering and characteristic path-length, interestingly, this model successfully realised number of FFMs as well as number of driver neurons. We found that for an intermediate distance constraint of $\alpha = 0.6$ this model is closest to CeNN in reproducing number of driver neurons that are critical for control of the network~(\figurename~\ref{fig:dcr_dcp}(d)). This model presents a range of distance constraint parameter $(0.6 \leq \alpha \leq 1.8)$ for which the small world nature as well as functionally relevant features of regulatory motifs and controllability were realistically exhibited. The number of driver nodes in DCP model was always higher than those returned by DCR model. For $\alpha < 1.8$, i.e.\ in the presence of strong distance constraint, DCP model returned significantly high number of driver nodes higher than maximally displayed by DCR model. This points at the role of long distance synaptic connections in conferring observed nature of control in CeNN. This result brings out the importance of optimum heterogeneity in the range at which synapses are formed. Economizing on the synaptic lengths yields the Cartesian model ($\alpha \rightarrow \infty$), equivalent of a 2D regular graph devoid of long-range connections. In such a case control nature of DCP model is similar to that of ER model. For smaller values of $\alpha$, where long range synaptic connections dominate, the model yields control response closest to reality. This highlights the important role played by long-range connections in controllability, and hence in key biological functions of the worm.

\subsection{Identification of specific driver neurons}

\begin{figure}[!t]
\centering
\includegraphics[scale=0.2]{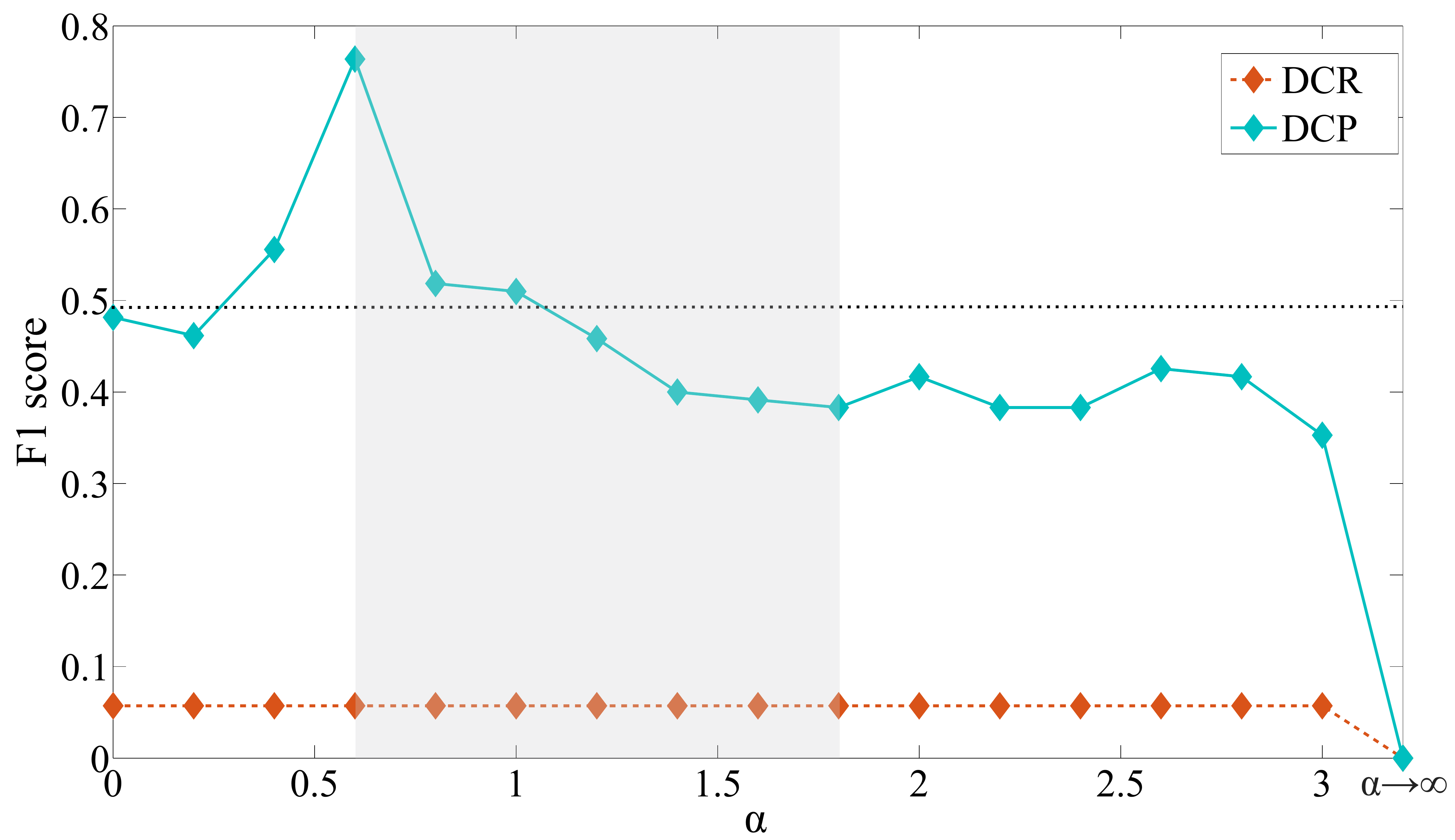}
\caption{Accuracy of identification of specific driver neurons with changing distance constraint exponent. The DCR model, having random synaptic connectivity pattern, fared poorly with accuracy comparable to random sampling. The performance of DCP model was consistently better than that of the DCR model indicating the critical role played by the neural connectivity and distance constraint in specifying the control of the neuronal network. For optimum distance constraint ($\alpha=0.6$) DCP model provides the best match with the reality ($F1~score = 0.77$), better than what could be accounted for by only neuronal connectivity (DD control; indicated with a dashed line). The spectrum of distance constraint regime for which DCP model is closest to CeNN ($0.6 \leq \alpha \leq 1.8$) is highlighted with gray background.}
\label{fig:specific_neurons}
\end{figure}

While the DCP model successfully reproduces key topological features important for function and control of CeNN~(\figurename~\ref{fig:dcr_dcp}), the question is whether it can also capture specific neurons implicated in control of the network and not just the number of driver neurons~(\figurename~\ref{fig:dcr_dcp}(d)). We compared specific neurons identified by `the minimum driver neurons set' obtained from real-world CeNN with that obtained from distance constraint models, for varying extent of distance constraint $\alpha$ 
~(\figurename~\ref{fig:specific_neurons}). Neurons that were consistently identified as driver neurons over 100 random instances of DCP and DCR models were obtained as True Positives. The classification accuracy of these models was assessed using F1 score. Interestingly, we found that the performance of DCP model was significantly better compared to DCR model (the accuracy of which was indistinguishable from that obtained from random sampling) and superior to DD control in the presence of dominant long distance connections ($\alpha \leq 1$). Thus DCP model is not only closer to real-world network in terms of number of driver nodes but also accurately identified specific neurons that can drive the network dynamics. In summary, the DCP model, that embeds empirically observed phenomenon of neuronal rewiring in addition to fixed neuronal connectivity, successfully recreates topological features of functional relevance to \textit{C.\ elegans}.

\section{Discussion}
\textit{C. elegans} connectome is one of the simplest yet complete neural diagram known to us so far. The neuronal network of this organism is responsible for many essential cognitive functions including learning and memory~\cite{Ardiel2010}. CeNN seemingly has evolved as a small world network with high clustering and low characteristic path-length for functional benefits~\cite{Watts1998}. Other than the small world global architecture CeNN is reported to be enriched with number of feed forward motifs among all possible three and four node motifs~\cite{Milo2002}. Our results suggest that the heterogeneous composition of motifs dictated by FFMs contributes to increased clustering as well as control of the network.

Analysis of neuronal architecture of CeNN has revealed that the network is optimally wired~\cite{Chen2006,Perez-Escudero2007} and is dictated by constraints~\cite{Ahn2006, Itzhack2010, Towlson2013, Pan2010}.
Till date a few simple null models of CeNN have been implemented with network feature constraints~\cite{Erdos1959, Maslov2002}. These studies suggest that neuronal connectivity plays a key role in rendering clustering as well as presentation of as many driver neurons as observed in CeNN~\cite{Liu2011}. None of these models has been able to explain all network features, especially clustering and number of driver neurons, claimed to be of biological relevance~\cite{Liu2011, Badhwar2015}.

Here, we present a distance constrained synaptic plasticity model that accounts for high clustering, FFMs saturation and large number of driver nodes. With a 1D ring model maximized for feed forward motifs, we show that such a model exhibits critical phenomenon in response to increased probability of synaptic rewiring. While this simple model lends interesting insights into the mechanisms of CeNN architecture, it cannot capture the aspect of controllability. This study indicated that small-world nature, known to be important for neuronal communication, can be recreated at an optimal value of rewiring. But at the same time 1D model is incapable of exhibiting nature of control observed in the empirical network.

Rooted in empirical observation of distance constraint followed in neuronal connections, we built more realistic 2D distance constrained models with random connectivity (DCR) and degree preserved connectivity (DCP). The latter model, that mimics real-world \textit{C. elegans} neuronal wiring and follows a distance constrained synaptic plasticity mechanism, comes closest to the CeNN in presenting small world architecture, dominance of FFMs and nature of controllability within a range of free variable $\alpha$. Consistent with previous studies, number of contacts made by neurons came out as an important factor that governs network control~\cite{Liu2011, Badhwar2015}. But importantly, this study brings out the key role played by long-range synaptic connections. Our results suggest that the extent of synaptic plasticity in CeNN is optimized so as to acquire key structural and dynamical network features. Thus, optimized long-range synaptic links in response to synaptic plasticity as a biologically relevant property that not only ensures small world nature, but also lends the control phenotype of this neuronal network. Further, the DCP model also successfully captures specific driver neurons with impressive accuracy. Thus beyond confirming the importance of connectivity (degree) of neurons and highlighting the role of long-range synapses as signalling conduits, this results identifies parameter value for which the model is closest to CeNN.

Structurally, the small world architecture and distance constrained connectivity mirror presence of densely connected ganglionic structures that are bridged via optimized long-range synapses~\cite{Ahn2006}. From functional perspective, optimum heterogeneity in synaptic lengths is a reflection of wiring optimization in the brain  architecture~\cite{Chen2006,Ahn2006,Perez-Escudero2007} of the worm that is shaped by evolution. The model reveals the role of length constrained wiring in shaping biologically important neuronal clustering~\cite{Ahn2006} and saturation of functionally critical feed forward motifs~\cite{Mangan2003, Milo2002}. It also pins down specific driver neurons that have been shown to be linked with critical functions of the worm such as reproduction, signalling processes and anatomical structural development~\cite{Badhwar2015}. Thus, the distance-constrained plasticity model presented in this study embodies an essential structural aspect of neuronal wiring with significant biological implications for survival of the organism.

Clearly, while the DCP model highlights the role of synaptic plasticity and distance constrained neuronal connectivity in specifying structural features and control architecture of CeNN, it is limited in many ways. The present study overlooks functional differences of synaptic links such as chemical synapses and gap junctions. Also, the strength of synaptic connections (edge weight) were ignored in these unweighted network models. Models that factor in such biologically relevant aspects, which are ignored in this study in favour of simplicity, may yield more enriched representations of CeNN.

\section*{Acknowledgment}
Ganesh Bagler acknowledges the seed grant support from Indian Institute of Technology Jodhpur (IITJ/SEED/2014/0003), and support from Dhirubhai Ambani Institute of Information and Communication Technology as well as from Indraprastha Institute of Information Technology Delhi (IIIT-Delhi). Rahul Badhwar thanks Ministry of Human Resource Development, Government of India and Indian Institute of Technology Jodhpur for the senior research fellowship.

%

\newpage
\onecolumngrid
\setcounter{table}{0}
\setcounter{figure}{0}
\setcounter{section}{0}

\renewcommand{\thetable}{S\arabic{table}}
\renewcommand{\thefigure}{S\arabic{figure}}
\renewcommand{\thesection}{S\arabic{section}}

\section*{Supplemental Material}
\section{Three node motif classification}
\begin{figure}[h]
\begin{center}
\includegraphics[scale=1]{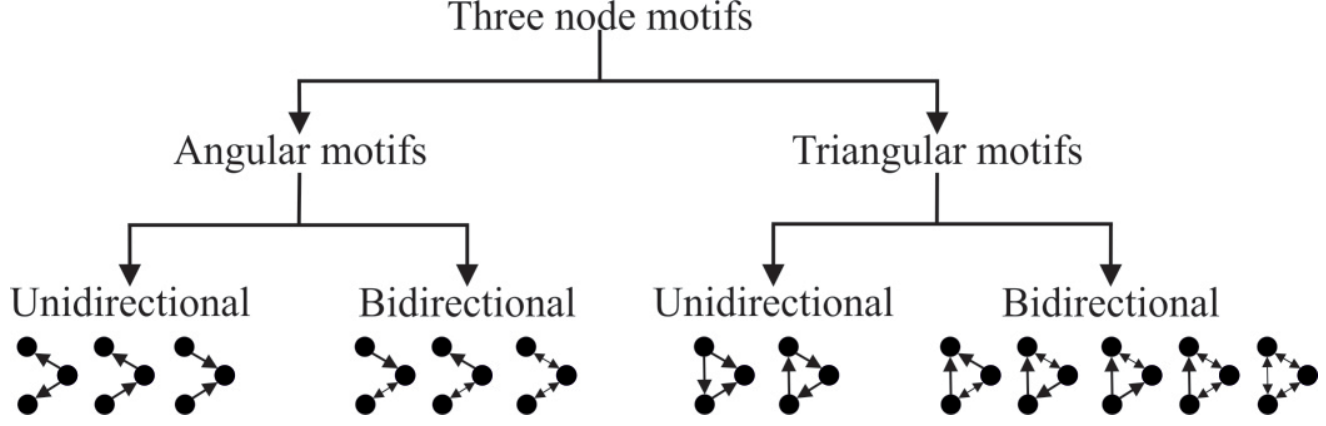}
\end{center}
\caption{\label{fig:motif_cat} Classification of three node subgraphs}
\end{figure}

\begin{figure}[h]
\begin{center}
\includegraphics[scale=0.27]{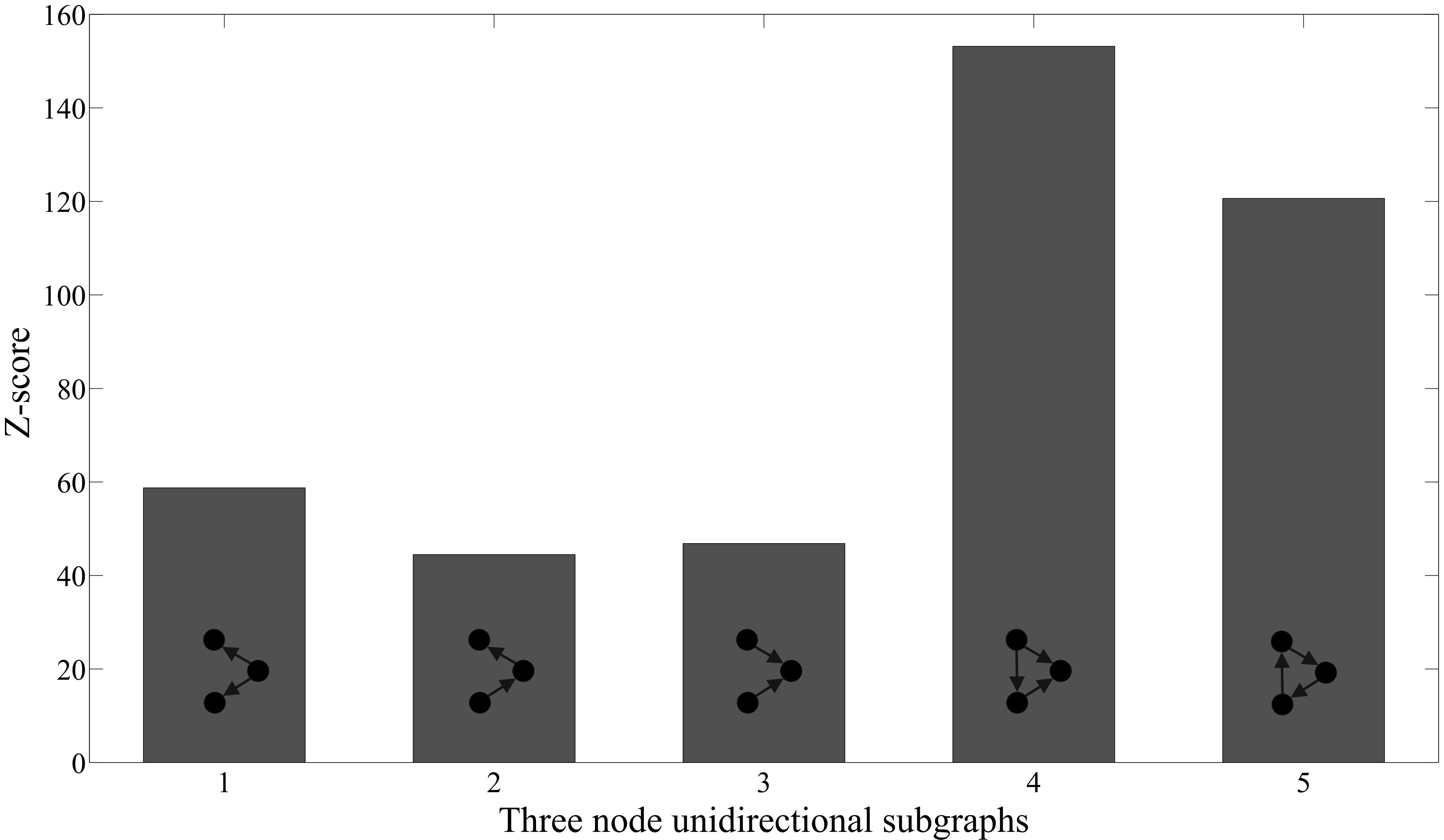}
\end{center}
\caption{\label{fig:motif_stats} Statistics for unidirectional three node motifs depicting over-representation of feed forward motifs. The Z-Score was computed in comparison to 100 instances of random controls (ER) of CeNN.}
\end{figure}

Motifs are patterns of local connectivity among nodes that are present in numbers significantly higher than expected by chance. The pattern of connectivity among 13 three node connected digraphs could be divided into angular motifs and triangular motifs~(Figure~\ref{fig:motif_cat}). Angular motifs are linear three node sub-structures, whereas triangular motifs comprise of three nodes subgraphs with either unidirectional or bidirectional edges. For our studies, we computed number of feed forward motifs, $n_{FFM}$, (among unidirectional triangular motifs) that are prevalent in many real world networks including CeNN. In CeNN, feed forward motifs are most prevalent among all unidirectional motifs as shown in Figure~\ref{fig:motif_stats}.

\newpage
\section{1D ring model of CeNN}
\begin{figure}[h]
\begin{center}
\includegraphics[scale=0.6]{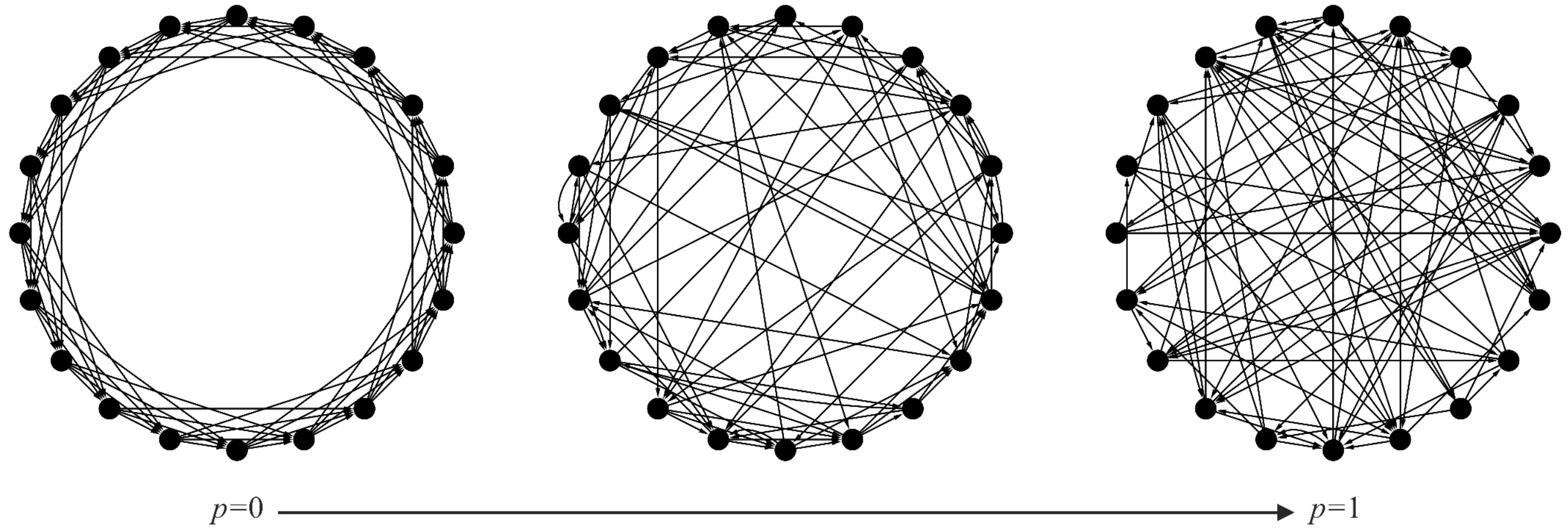}
\end{center}
\caption{\label{fig:Ring1D_20nodes} Regular graph for maximum number of Feed forward motifs with graph growth. Edges are directed and point from nodes to nearest neighbours along with neighbour of neighbours and so on. This representation if for $n = 20$ and $k = 4$.}
\end{figure}

Following is an illustration of 1D ring model with 20 nodes and 4 outgoing synapses for every node($n = 20$ and $k = 4$)~(Figure~\ref{fig:Ring1D_20nodes}).

\section{Response of CeNN 1D ring model to rewiring}
\begin{figure}[H]
\begin{center}
\includegraphics[scale=0.3]{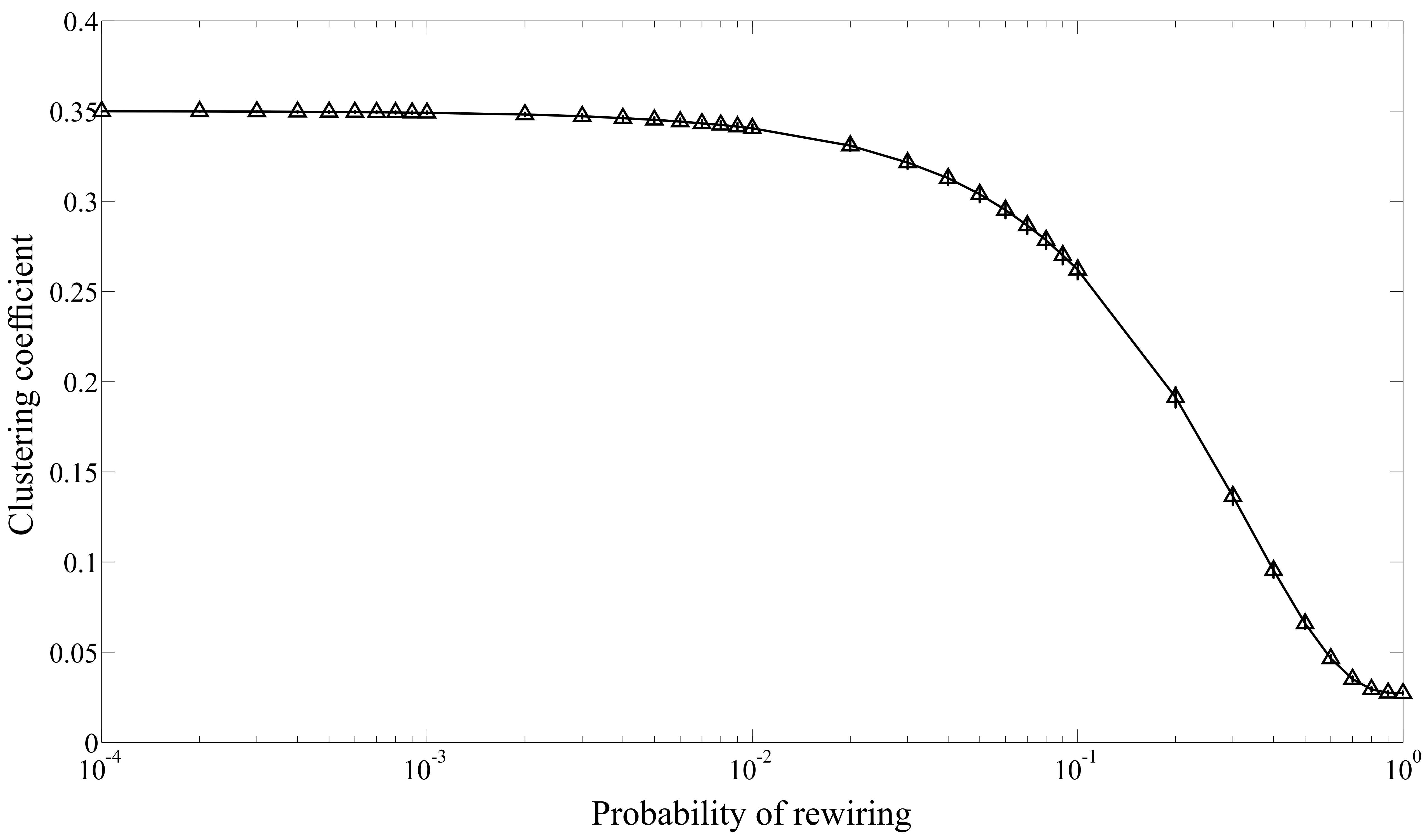}
\end{center}
\caption{\label{fig:Stats_1Drewire_C} Response of average clustering coefficient to rewiring.}
\end{figure}

\begin{figure}[H]
\begin{center}
\includegraphics[scale=0.3]{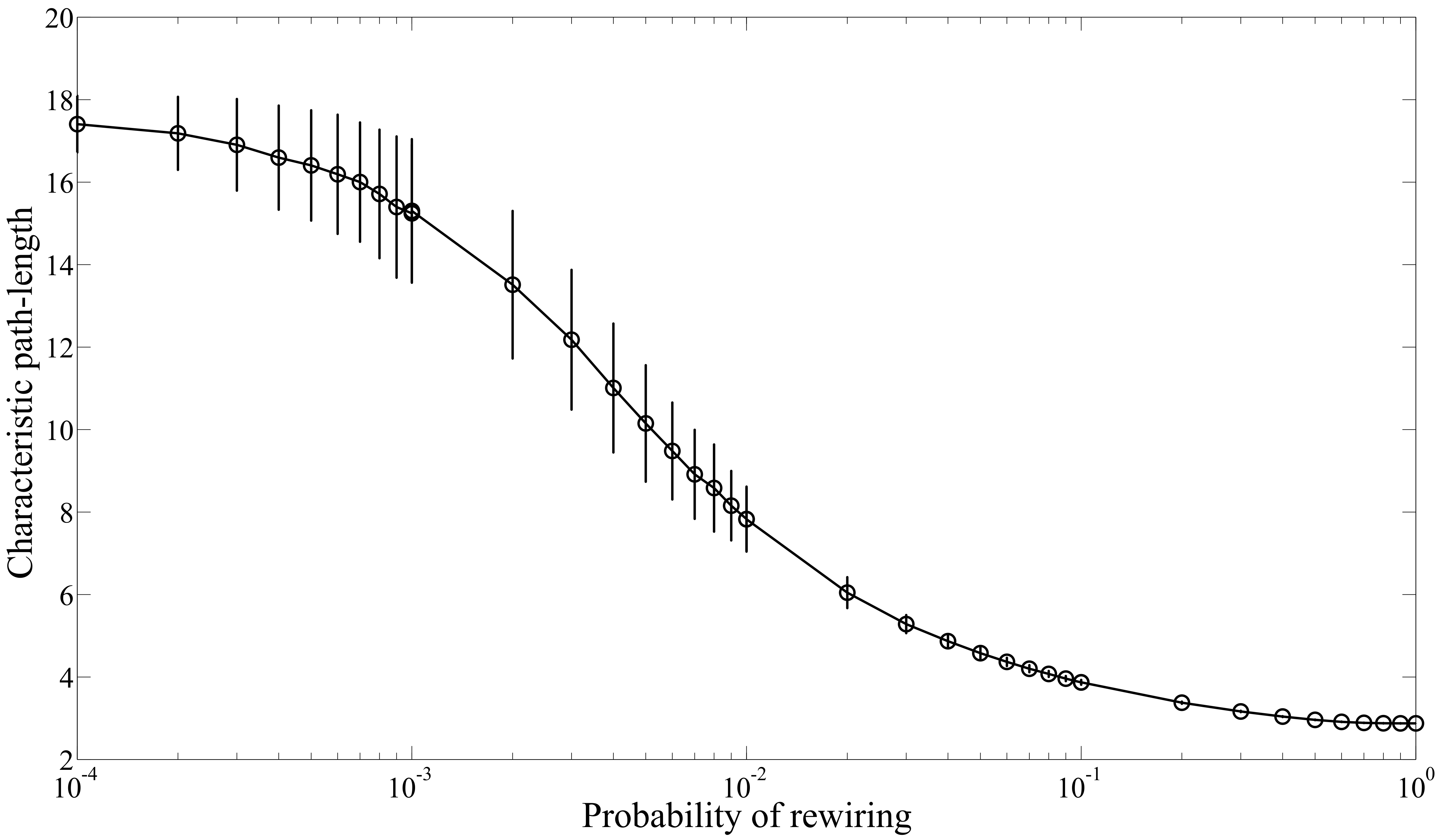}
\end{center}
\caption{\label{fig:Stats_1Drewire_L} Response of characteristic path-length to rewiring.}
\end{figure}

\begin{figure}[H]
\begin{center}
\includegraphics[scale=0.3]{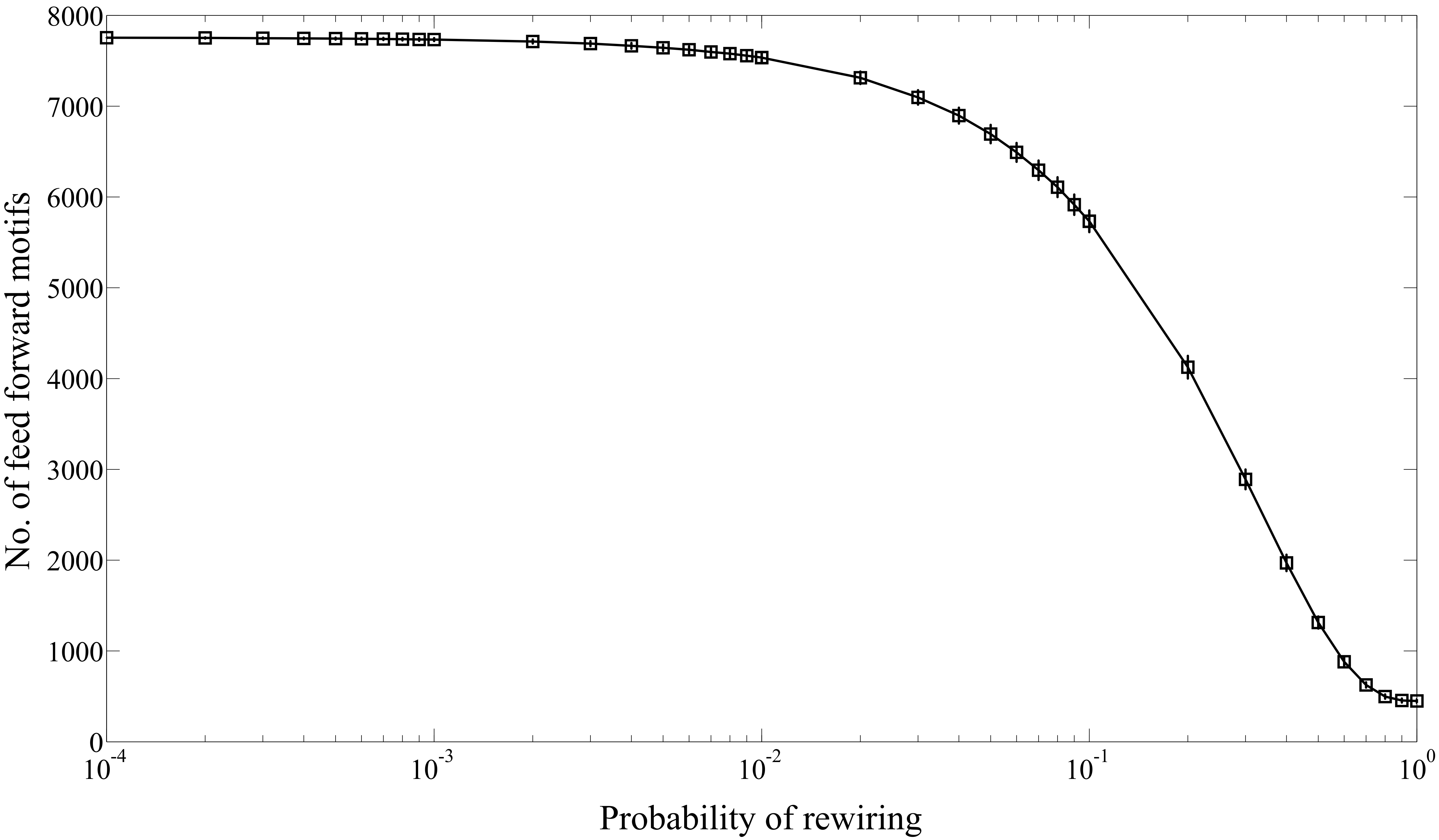}
\end{center}
\caption{\label{fig:Stats_1Drewire_Nffm} Response of number of feed forward motifs to rewiring.}
\end{figure}

Starting from a 1-D regular ring graph maximally saturated with $7756$ FFMs, we simulated random synaptic rewiring to observe its effect on topological features. In addition FFMs saturation, the regular graph had very high average clustering coefficient ($\overline{C}_{reg} = 0.35$, Figure~\ref{fig:Stats_1Drewire_C}) as well as characteristic path-length ($L_{reg} = 17.69$; Figure~\ref{fig:Stats_1Drewire_L}). Figure~\ref{fig:Stats_1Drewire_Nffm} shows with increasing probability of synaptic rewiring the number of FFMs is unaffected up to $p\approx 0.01$ before falling sharply. These figures depict non-normalized data corresponding to Figure~3 in the main manuscript.

\newpage
\section{Degree distribution of CeNN, its controls and distance constrained models}
\begin{figure}[h]
\begin{center}
\includegraphics[scale=0.8]{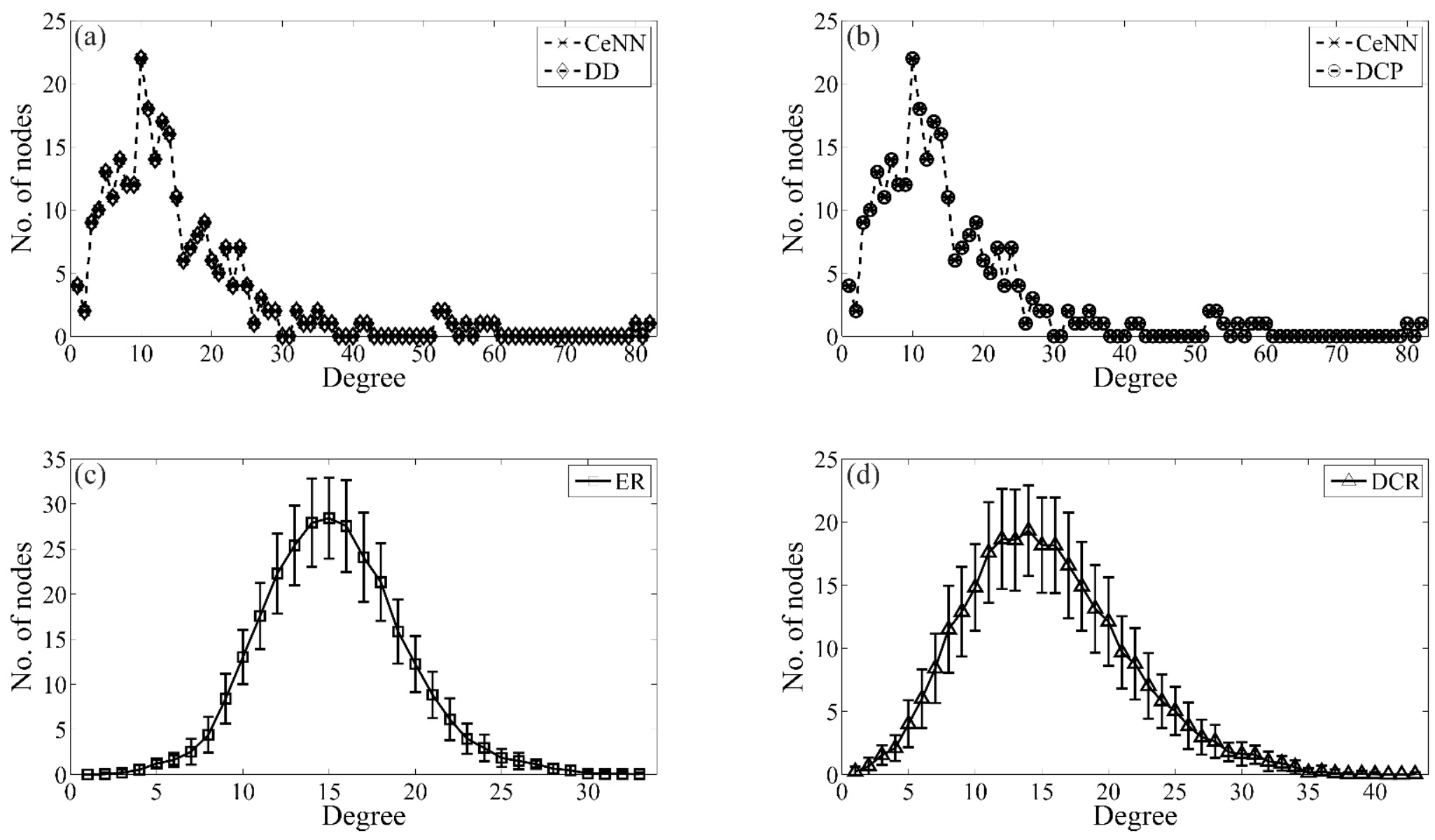}
\end{center}
\caption{\label{fig:degree_dist} Degree distributions of CeNN, its controls and distance constrained models.}
\end{figure}

Figure~\ref{fig:degree_dist} depicts the degree distribution of CeNN in comparison to its random controls (ER and DD) as well as distance constrained models (DCR and DCP).

\newpage
\section{Topological features of DCR}

\begin{table}[h]
\centering
\caption{\label{Tab:DCR} Differential topological properties of distance constraint random model at different values of $\alpha$ starting from an ER control.}

\begin{center}
\begin{tabular}{|l|l|l|l|l|l|l|l|l|}
\hline
 & Average & Characteristic & Number of & Number of \\
Exponent & clustering coefficient $(\overline{C})$ & path-length $(L)$ & driver neurons $(n_{D})$ & feed forward motifs \\
 \hline
$\alpha = 0$ (ER) &0.032 $\pm$ 0.001 &2.968 $\pm$ 0.007 &0.26 $\pm$ 0.441 &610.55 $\pm$ 37.59 \\
\hline
$\alpha = 0.2$ &0.032 $\pm$ 0.002 &3.048 $\pm$ 0.026 &13.75 $\pm$ 2.536 &563.91 $\pm$ 29.212\\
\hline
$\alpha = 0.4$ &0.033 $\pm$ 0.001 &3.068 $\pm$ 0.026 &15.70 $\pm$ 2.866 &612.19 $\pm$ 33.902 \\
\hline
$\alpha = 0.6$ &0.039 $\pm$ 0.002 &3.14 $\pm$ 0.028 &15.24 $\pm$ 2.934 &721.21 $\pm$ 38.011 \\
\hline
$\alpha = 0.8$ &0.048 $\pm$ 0.002 &3.211 $\pm$ 0.042 &14.66 $\pm$ 2.952 &875.83 $\pm$ 40.081 \\
\hline
$\alpha = 1.0$ &0.064 $\pm$ 0.003 &3.334 $\pm$ 0.044 &13.62 $\pm$ 2.784 &1131.84 $\pm$ 53.844 \\
\hline
$\alpha = 1.2$ &0.085 $\pm$ 0.004 &3.49 $\pm$ 0.07 &11.82 $\pm$ 2.418 &1447.24 $\pm$ 65.112 \\
\hline
$\alpha = 1.4$ &0.108 $\pm$ 0.004 &3.689 $\pm$ 0.099 &9.37 $\pm$ 2.299 & 1816.30 $\pm$ 73.181 \\
\hline
$\alpha = 1.6$ &0.136 $\pm$ 0.005 &3.931 $\pm$ 0.14 &8.08 $\pm$ 2.246 &2231.32 $\pm$ 87.442 \\
\hline
$\alpha = 1.8$ &0.169 $\pm$ 0.006 &4.185 $\pm$ 0.266 &7.23 $\pm$ 2.287 &2719.38 $\pm$ 97.224 \\
\hline
$\alpha = 2.0$ &0.196 $\pm$ 0.007 &4.426 $\pm$ 0.408 &5.73 $\pm$ 1.89 &3114.38 $\pm$ 108.00 \\
\hline
$\alpha = 2.2$ &0.227 $\pm$ 0.008 &4.589 $\pm$ 0.617 &5.09 $\pm$ 1.682 &3562.71 $\pm$ 121.276 \\
\hline
$\alpha = 2.4$ &0.256 $\pm$ 0.008 &4.744 $\pm$ 0.661 &4.98 $\pm$ 1.826 &3985.73 $\pm$ 121.322 \\
\hline
$\alpha = 2.6$ &0.28 $\pm$ 0.008 &4.726 $\pm$ 0.709 &4.46 $\pm$ 1.72 &4364.37 $\pm$ 118.252 \\
\hline
$\alpha = 2.8$ &0.305 $\pm$ 0.008 &4.976 $\pm$ 0.843 &4.17 $\pm$ 1.596 &4738.90 $\pm$ 124.48 \\
\hline
$\alpha = 3.0$ &0.326 $\pm$ 0.008 &5.138 $\pm$ 0.856 &3.93 $\pm$ 1.725 &5035.94 $\pm$ 120.082 \\
\hline
$\alpha \rightarrow \infty$ & & & & \\
Cartesian & 0.624 & 10.394 & 0 & 10153 \\
\hline
\end{tabular}
\end{center}

\end{table}

\section{Topological features of DCP}

\begin{center}
\begin{table}[h]
 \centering
\caption{\label{Tab:DCP} Differential properties of distance constraint synaptic plasticity model at different values of $\alpha$ starting from DD control. }

\begin{tabular}{|l|l|l|l|l|l|l|l|l|}\hline
 & Average & Characteristic & Number of & Number of \\
 Exponent & clustering coefficient $(\overline{C})$ & pathlength $(L)$ & driver neurons $(n_{D})$ & feed forward motifs \\\hline
$\alpha = 0$ (DD) &0.067 $\pm$ 0.003 &2.98 $\pm$ 0.018 &22.38 $\pm$ 1.153 &1699.56 $\pm$ 57.496 \\\hline
$\alpha = 0.2$ &0.047 $\pm$ 0.002 &3.116 $\pm$ 0.024 &20.37 $\pm$ 2.427 &1155.70 $\pm$ 51.703 \\\hline
$\alpha = 0.4$ &0.051 $\pm$ 0.002 &3.133 $\pm$ 0.034 &22.37 $\pm$ 2.863 &1379.11 $\pm$ 61.456 \\\hline
$\alpha = 0.6$ &0.058 $\pm$ 0.003 &3.182 $\pm$ 0.033 &23.66 $\pm$ 2.503 &1667.59 $\pm$ 66.845 \\\hline
$\alpha = 0.8$ &0.069 $\pm$ 0.003 &3.286 $\pm$ 0.041 &23.07 $\pm$ 2.808 &2026.05 $\pm$ 72.395 \\\hline
$\alpha = 1.0$ &0.086 $\pm$ 0.004 &3.423 $\pm$ 0.055 &21.51 $\pm$ 3.422 &2402.08 $\pm$ 97.428 \\\hline
$\alpha = 1.2$ &0.107 $\pm$ 0.005 &3.58 $\pm$ 0.075 &19.70 $\pm$ 2.333 &2832.55 $\pm$ 89.72 \\\hline
$\alpha = 1.4$ &0.131 $\pm$ 0.005 &3.79 $\pm$ 0.112 &17.62 $\pm$ 2.112 &3263.43 $\pm$ 106.606 \\\hline
$\alpha = 1.6$ &0.158 $\pm$ 0.006 &4.063 $\pm$ 0.16 &16.36 $\pm$ 1.784 &3726.71 $\pm$ 109.069  \\\hline
$\alpha = 1.8$ &0.186 $\pm$ 0.006 &4.384 $\pm$ 0.317 &15.89 $\pm$ 1.377 &4150.42 $\pm$ 106.657 \\\hline
$\alpha = 2.0$ &0.215 $\pm$ 0.007 &4.615 $\pm$ 0.439 &15.67 $\pm$ 1.28 &4597.74 $\pm$ 114.656  \\\hline
$\alpha = 2.2$ &0.241 $\pm$ 0.007 &4.829 $\pm$ 0.576 &15.55 $\pm$ 1.167 &5025.00 $\pm$ 120.231 \\\hline
$\alpha = 2.4$ &0.266 $\pm$ 0.008 &4.947 $\pm$ 0.702 &15.28 $\pm$ 1.092 &5391.46 $\pm$ 114.416 \\\hline
$\alpha = 2.6$ &0.288 $\pm$ 0.009 &5.256 $\pm$ 0.902 &15.05 $\pm$ 0.947 &5722.95 $\pm$ 121.129 \\\hline
$\alpha = 2.8$ &0.311 $\pm$ 0.008 &5.256 $\pm$ 1.046 &14.95 $\pm$ 0.925 &6084.32 $\pm$ 118.954 \\\hline
$\alpha = 3.0$ &0.328 $\pm$ 0.009 &5.395 $\pm$ 1.002 &15.10 $\pm$ 0.959 &6339.36 $\pm$ 131.354 \\\hline
$\alpha \rightarrow \infty$ & & & & \\
Cartesian & 0.624 & 10.394 & 0 & 10153 \\\hline
\end{tabular}
\end{table}
\end{center}

\section{List of \emph{C.\ elegans} neurons}
Table~\ref{Tab:neuronsXY} lists all 277 neurons with their X- and Y-coordiantes. Rows depicting the 34 driver neurons identified in this study are shown in grey.

\begin{center}
\begin{table}[h]
\centering
\caption{\label{Tab:neuronsXY} Driver neurons of CeNN.}
\begin{tabular}{ | l | l | l |}
\hline
	\textbf{Neuron Name} & \textbf{X} & \textbf{Y} \\ \hline
	ADAL & 0.01106776858176 & 0.00590280993408  \\  \hline
	ADAR & 0.01420641977856 & 0.00220444444224  \\ \hline
	ADEL & 0.01623272728896 & 0.00565685948544  \\ \hline
	ADER & 0.01494123459264 & 0.00930765433152  \\ \hline
	ADFL & 0.08239338844992 & -0.00098380167552  \\ \hline
	ADFR & 0.0832790123616 & -0.00318419750784  \\ \hline
	ADLL & 0.08263933883904 & -0.01303537187328  \\ \hline
	ADLR & 0.0832790123616 & -0.01151209877376  \\ \hline
	AFDL & 0.08632859503296 & -0.0027054545184  \\ \hline
	AFDR & 0.08646320986944 & -0.0009797530656  \\ \hline
	AIAL & 0.06517685948544 & 0.00934611567936  \\ \hline
	AIAR & 0.05903012343744 & 0.01151209877376  \\ \hline
	AIBR & 0.07544098765824 & 0.00612345676416  \\ \hline
	AIML & 0.0332033058048 & 0.01795438019136  \\ \hline
	AIMR & 0.0374755555776 & 0.01518617284416  \\ \hline
	AINL & 0.06197950412928 & -0.0061487603232  \\ \hline
	AINR & 0.06196938269376 & -0.00342913581888  \\ \hline
	AIYR & 0.0394350617088 & 0.01518617284416  \\ \hline
	AIZL & 0.04869818180736 & 0.0027054545184  \\ \hline
	AIZR & 0.05756049380928 & 0.00318419750784  \\ \hline
	ALA & 0.09405629632128 & -0.01322666665344  \\ \hline
	ALML & -0.37848615384768 & -0.01678769230464  \\ \hline
	ALMR & -0.3815384615232 & -0.00610461541056  \\ \hline
	ALNL & -0.9800451284544 & -0.04944102563136  \\ \hline
	ALNR & -0.9862451280576 & -0.04944102563136  \\ \hline
	AQR & 0.01959506172864 & 0.00416395063296  \\ \hline
	AS1 & -0.0164108642208 & 0.0247387654272  \\ \hline
	AS10 & -0.8348061541056 & 0.03357538460928  \\ \hline
	AS11 & -0.8440584613248 & 0.02330564102208  \\ \hline
	AS2 & -0.10988307691392 & 0.03968000001984  \\ \hline
	AS3 & -0.19839999998016 & 0.03968000001984  \\ \hline
	AS4 & -0.28386461537088 & 0.03968000001984  \\ \hline
	AS5 & -0.40137846156288 & 0.04273230769536  \\ \hline
	AS6 & -0.47616 & 0.04273230769536  \\ \hline
	AS7 & -0.57536000001984 & 0.04273230769536  \\ \hline
	AS8 & -0.6654030769536 & 0.03968000001984  \\ \hline
	AS9 & -0.7584984613248 & 0.0351015384768  \\ \hline
	ASEL & 0.069112066128 & 0.00049190083776  \\ \hline
	ASER & 0.07176691358784 & -0.00073481481408  \\ \hline
	ASGL & 0.0779662809696 & -0.00762446283648  \\ \hline
	ASGR & 0.08009481479424 & -0.01028740739712  \\ \hline
	ASHL & 0.07550677684032 & 0.00049190083776  \\ \hline
	ASHR & 0.07838024691456 & -0.0009797530656  \\ \hline
	ASIL & 0.0769824793536 & -0.01278942148416  \\ \hline
	ASIR & 0.07715555553792 & -0.01175703702528  \\ \hline
	ASJL & 0.06345520658304 & 0.00664066116096  \\ \hline
	ASJR & 0.06270419750784 & 0.00906271602048  \\ \hline
	ASKL & 0.08854214877312 & -0.01008396696576  \\ \hline
	ASKR & 0.08866765431168 & -0.00955259258304  \\ \hline
\end{tabular}
\end{table}
\end{center}

\begin{center}
\begin{table}[h]
\centering
\begin{tabular}{ | l | l | l | }
\hline
	\textbf{Neuron Name} & \textbf{X} & \textbf{Y} \\ \hline
	AUAL & 0.06862016529024 & 0.0061487603232  \\ \hline
	AUAR & 0.06711308639232 & 0.00538864195008  \\ \hline
	AVAL & 0.08952595038912 & 0.0017216529024  \\ \hline
	AVAR & 0.09087209875392 & -0.00024493825152  \\ \hline
	AVBL & 0.069112066128 & -0.0044271074208  \\ \hline
	AVBR & 0.07176691358784 & -0.0063683950752  \\ \hline
	AVDL & 0.06173355374016 & -0.00147570245376  \\ \hline
	AVDR & 0.06637827157824 & -0.00146962962816  \\ \hline
	AVEL & 0.08288528922816 & 0.00221355374016  \\ \hline
	AVER & 0.08401382717568 & 0.00318419750784  \\ \hline
	AVFL & 0.01861530866304 & 0.02277925923648  \\ \hline
	AVFR & 0.02449382717568 & 0.02253432098496  \\ \hline
	AVG & 0.00195950619072 & 0.02449382717568  \\ \hline
	AVHL & 0.07206347109504 & -0.00860826445248  \\ \hline
	AVHR & 0.07642074072384 & -0.0127367901504  \\ \hline
	AVJL & 0.06763636361472 & -0.00934611567936  \\ \hline
	AVJR & 0.07299160496448 & -0.00979753083456  \\ \hline
	AVKL & 0.02680859503296 & 0.01770842974272  \\ \hline
	AVKR & 0.03184197531648 & 0.01592098765824  \\ \hline
	AVL & 0.06098962962816 & 0.00906271602048  \\ \hline
	AVM & -0.34796307691392 & 0.0228923077152  \\ \hline
	AWAL & 0.0779662809696 & -0.00295140496704  \\ \hline
	AWAR & 0.07838024691456 & -0.00538864195008  \\ \hline
	AWBL & 0.07919603303424 & -0.00393520658304  \\ \hline
	AWBR & 0.08278913579904 & -0.00685827157824  \\ \hline
	AWCL & 0.07845818180736 & 0.00491900825856  \\ \hline
	AWCR & 0.07862518516608 & 0.00416395063296  \\ \hline
	BAGL & 0.1128912396768 & 0.00049190083776  \\ \hline
	BAGR & 0.11463111111552 & 0.00318419750784  \\ \hline
	BDUL & -0.13277538462912 & -0.00152615386752  \\ \hline
	BDUR & -0.13430153843712 & 0.00457846154304  \\ \hline
	CEPDL & 0.09419900825856 & -0.01623272728896  \\ \hline
	CEPDR & 0.09601580245248 & -0.01249185183936  \\ \hline
	CEPVL & 0.10748033058048 & 0.0113137190304  \\ \hline
	CEPVR & 0.10507851853248 & 0.01494123459264  \\ \hline
	DA1 & -0.02302419754752 & 0.02424888886464  \\ \hline
	DA2 & -0.12819692308608 & 0.03815384615232  \\ \hline
	DA3 & -0.23350153845696 & 0.04120615382784  \\ \hline
	DA4 & -0.37390769230464 & 0.04120615382784  \\ \hline
	DA5 & -0.52652307691392 & 0.03968000001984  \\ \hline
	DA6 & -0.6989784613248 & 0.03815384615232  \\ \hline
	DA7 & -0.8409107693376 & 0.03204923074176  \\ \hline
	DA8 & -0.9039917949888 & 0.0038789743488  \\ \hline
	DA9 & -0.9077117949888 & 0.00263897436864  \\ \hline
	DB1 & -0.01224691358784 & 0.02522864198976  \\ \hline
	DB2 & 0.01175703702528 & 0.02424888886464  \\ \hline
	DB3 & -0.12056615386752 & 0.03815384615232  \\ \hline
	DB4 & -0.2731815384768 & 0.03968000001984  \\ \hline
\end{tabular}
\end{table}
\end{center}

\begin{center}
\begin{table}[h]
\centering
\begin{tabular}{ | l | l | l |}
\hline
	\textbf{Neuron Name} & \textbf{X} & \textbf{Y} \\ \hline
	DB5 & -0.46852923078144 & 0.04120615382784  \\ \hline
	DB6 & -0.657772307616 & 0.03815384615232  \\ \hline
	DB7 & -0.827175384768 & 0.03204923074176  \\ \hline
	DD1 & -0.00808296295488 & 0.0247387654272  \\ \hline
	DD2 & -0.18924307689408 & 0.03815384615232  \\ \hline
	DD3 & -0.37848615384768 & 0.04120615382784  \\ \hline
	DD4 & -0.5829907692384 & 0.04120615382784  \\ \hline
	DD5 & -0.767655384768 & 0.0366276922848  \\ \hline
	DD6 & -0.895311794592 & 0.00511897438848  \\ \hline
	DVA & -0.9444984613248 & -0.03786769232448  \\ \hline
	DVB & -0.939951794592 & -0.03580102561152  \\ \hline
	DVC & -0.951525128256 & -0.03786769232448  \\ \hline
	FLPL & 0.02336528922816 & 0.00910016529024  \\ \hline
	FLPR & 0.02498370367872 & 0.00391901232192  \\ \hline
	HSNL & -0.53873230767552 & 0.00915692308608  \\ \hline
	HSNR & -0.54178461541056 & 0.0351015384768  \\ \hline
	IL1DL & 0.12814016529024 & -0.01451107438656  \\ \hline
	IL1DR & 0.12687802470336 & -0.0100424691456  \\ \hline
	IL1L & 0.12641851238784 & -0.00098380167552  \\ \hline
	IL1R & 0.12418370369856 & 0.00244938269376  \\ \hline
	IL1VL & 0.12912396696576 & 0.01180561980864  \\ \hline
	IL1VR & 0.11218172842176 & 0.01322666665344  \\ \hline
	IL2DL & 0.13674842974272 & -0.0157408264512  \\ \hline
	IL2DR & 0.137410370352 & -0.01396148146752  \\ \hline
	IL2L & 0.13158347109504 & -0.00295140496704  \\ \hline
	IL2R & 0.13030716052224 & -0.00073481481408  \\ \hline
	IL2VL & 0.13404297522432 & 0.01303537187328  \\ \hline
	IL2VR & 0.12687802470336 & 0.01224691358784  \\ \hline
	LUAL & -0.9730184615232 & -0.02050769230464  \\ \hline
	LUAR & -0.9833517947904 & -0.0180276922848  \\ \hline
	OLLL & 0.128632066128 & -0.0061487603232  \\ \hline
	OLLR & 0.1290824691456 & -0.00440888888448  \\ \hline
	OLQDL & 0.12272925619392 & -0.01598677684032  \\ \hline
	OLQDR & 0.1217343210048 & -0.01224691358784  \\ \hline
	OLQVL & 0.11608859503296 & 0.00418115703168  \\ \hline
	OLQVR & 0.114386172864 & 0.00759308639232  \\ \hline
	PDA & -0.9300317949888 & -0.00025435895808  \\ \hline
	\rowcolor{lightgray} PDB & -0.9130851284544 & 0.00305230767552  \\ \hline
	PDEL & -0.6821907691392 & -0.0122092307616  \\ \hline
	PDER & -0.6928738460928 & 0.02136615384768  \\ \hline
	PHAL & -0.958551794592 & -0.01596102563136  \\ \hline
	PHAR & -0.9606184617216 & -0.01678769230464  \\ \hline
	PHBL & -0.980871794592 & -0.01513435895808  \\ \hline
	PHBR & -0.9767384615232 & -0.0180276922848  \\ \hline
	PHCL & -1.0172451284544 & -0.01637435899776  \\ \hline
	PHCR & -1.0139384615232 & -0.02381435899776  \\ \hline
	PLML & -1.0250984615232 & -0.0184410256512  \\ \hline
	PLMR & -1.0230317949888 & -0.01720102561152  \\ \hline
	
\end{tabular}
\end{table}
\end{center}

\begin{center}
\begin{table}[h]
\centering
\begin{tabular}{ | l | l | l |}
\hline
	\textbf{Neuron Name} & \textbf{X} & \textbf{Y} \\ \hline
	PLNL & -1.0085651280576 & -0.0143076922848  \\ \hline
	PLNR & -0.970125128256 & -0.01761435897792  \\ \hline
	PQR & -0.9903784617216 & -0.0143076922848  \\ \hline
	PVCL & -0.9767384615232 & -0.0403476922848  \\ \hline
	PVCR & -0.9899651280576 & -0.03042769232448  \\ \hline
	PVDL & -0.6959261537088 & -0.0122092307616  \\ \hline
	PVDR & -0.6989784613248 & 0.01831384617216  \\ \hline
	PVM & -0.7035569230464 & -0.00915692308608  \\ \hline
	PVNL & -1.0242717947904 & -0.03249435897792  \\ \hline
	PVNR & -1.018071794592 & -0.0333210256512  \\ \hline
	PVPL & -0.895725128256 & 0.00966564100224  \\ \hline
	PVPR & -0.8783651280576 & 0.01338564100224  \\ \hline
	PVQL & -0.9630984613248 & -0.01926769232448  \\ \hline
	PVQR & -0.9531784617216 & -0.02422769230464  \\ \hline
	PVR & -0.9945117947904 & -0.03869435899776  \\ \hline
	\rowcolor{lightgray} PVT & -0.8824984617216 & 0.0150389743488  \\ \hline
	PVWL & -0.9990584615232 & -0.0180276922848  \\ \hline
	PVWR & -1.0077384613248 & -0.03621435897792  \\ \hline
	RIAL & 0.08952595038912 & -0.00393520658304  \\ \hline
	RIAR & 0.08989234568832 & -0.004653827136  \\ \hline
	RIBL & 0.07034181819264 & 0.0044271074208  \\ \hline
	RIBR & 0.07103209877376 & 0.00514370369856  \\ \hline
	RICL & 0.05583074380608 & 0.0061487603232  \\ \hline
	RICR & 0.05290666667328 & 0.00489876544704  \\ \hline
	RID & 0.09871012345728 & -0.0164108642208  \\ \hline
	RIFL & 0.00440888888448 & 0.02449382717568  \\ \hline
	RIFR & 0.01665580247232 & 0.02302419754752  \\ \hline
	RIGL & -0.0063683950752 & 0.0247387654272  \\ \hline
	RIGR & 0.0083279012064 & 0.02302419754752  \\ \hline
	RIH & 0.07993388432064 & 0.009592066128  \\ \hline
	RIML & 0.06443900825856 & 0.00245950412928  \\ \hline
	RIMR & 0.0668681481408 & 0.00342913581888  \\ \hline
	RIPL & 0.12149950412928 & -0.00664066116096  \\ \hline
	RIPR & 0.12099950619072 & -0.00318419750784  \\ \hline
	RIR & 0.0695624691456 & 0.00930765433152  \\ \hline
	RIS & 0.03453629632128 & 0.01616592590976  \\ \hline
	RIVL & 0.07304727271104 & -0.01820033058048  \\ \hline
	RIVR & 0.07617580247232 & -0.01690074072384  \\ \hline
	RMDDL & 0.07895008264512 & 0.00934611567936  \\ \hline
	RMDDR & 0.07715555553792 & 0.01224691358784  \\ \hline
	RMDL & 0.08731239670848 & 0.0051649587072  \\ \hline
	RMDR & 0.0879328394976 & 0.00514370369856  \\ \hline
	RMDVL & 0.09419900825856 & -0.00295140496704  \\ \hline
	RMDVR & 0.09724049382912 & -0.00195950619072  \\ \hline
	RMED & 0.1155966941952 & -0.01795438019136  \\ \hline
	\rowcolor{lightgray} RMEL & 0.11264528922816 & -0.00295140496704  \\ \hline
	\rowcolor{lightgray} RMER & 0.11291654323584 & -0.00024493825152  \\ \hline
	RMEV & 0.09405629632128 & 0.0173906172864  \\ \hline
	
\end{tabular}
\end{table}
\end{center}
	
\begin{center}
\begin{table}[h]
\centering
\begin{tabular}{ | l | l | l |}
\hline
	\textbf{Neuron Name} & \textbf{X} & \textbf{Y} \\ \hline
	RMFL & 0.069112066128 & 0.01278942148416  \\ \hline
	RMFR & 0.07127703702528 & 0.01224691358784  \\ \hline
	RMGL & 0.00885421490112 & 0.0034433058048  \\ \hline
	RMGR & 0.00955259258304 & 0.00171456787968  \\ \hline
	RMHL & 0.07304727271104 & 0.00910016529024  \\ \hline
	RMHR & 0.0722567901504 & 0.0100424691456  \\ \hline
	SAADL & 0.07772033058048 & 0.00934611567936  \\ \hline
	SAADR & 0.07838024691456 & 0.01151209877376  \\ \hline
	SAAVL & 0.0944449587072 & -0.00590280993408  \\ \hline
	SAAVR & 0.09454617282432 & -0.0026943210048  \\ \hline
	SABD & -0.0026943210048 & 0.02424888886464  \\ \hline
	\rowcolor{lightgray} SABVL & 0.0247387654272 & 0.02228938273344  \\ \hline
	\rowcolor{lightgray} SABVR & 0.02032987654272 & 0.02277925923648  \\ \hline
	SDQL & -0.6898215384768 & -0.01068307689408  \\ \hline
	SDQR & -0.14651076925824 & -0.00305230767552  \\ \hline
	\rowcolor{lightgray} SIADL & 0.06837421490112 & 0.01082181819264  \\ \hline
	\rowcolor{lightgray} SIADR & 0.06147950619072 & 0.01494123459264  \\ \hline
	\rowcolor{lightgray} SIAVL & 0.05189553716352 & 0.0140191735488  \\ \hline
	\rowcolor{lightgray} SIAVR & 0.05462123455296 & 0.01126716052224  \\ \hline
	\rowcolor{lightgray} SIBDL & 0.079687933872 & 0.00688661155008  \\ \hline
	\rowcolor{lightgray} SIBDR & 0.07984987654272 & 0.0073481481408  \\ \hline
	\rowcolor{lightgray} SIBVL & 0.07427702477568 & 0.01008396696576  \\ \hline
	\rowcolor{lightgray} SIBVR & 0.073236543216 & 0.00979753083456  \\ \hline
	SMBDL & 0.0646849587072 & 0.0140191735488  \\ \hline
	SMBDR & 0.05731555555776 & 0.01690074072384  \\ \hline
	SMBVL & 0.05952 & 0.01426512393792  \\ \hline
	SMBVR & 0.06613333332672 & 0.0127367901504  \\ \hline
	SMDDL & 0.07255537187328 & 0.01328132232192  \\ \hline
	SMDDR & 0.07078716052224 & 0.0146962962816  \\ \hline
	SMDVR & 0.09381135801024 & -0.00489876544704  \\ \hline
	URADL & 0.1313375206464 & -0.01205157025728  \\ \hline
	URADR & 0.12491851851264 & -0.0083279012064  \\ \hline
	URAVL & 0.11658049587072 & 0.00934611567936  \\ \hline
	URAVR & 0.11340641973888 & 0.01126716052224  \\ \hline
	URBL & 0.12100760329152 & -0.00295140496704  \\ \hline
	URBR & 0.12222419750784 & -0.00122469137664  \\ \hline
	URXL & 0.09223140496704 & -0.01180561980864  \\ \hline
	URXR & 0.09307654319616 & -0.0100424691456  \\ \hline
	URYDL & 0.12568066116096 & -0.01205157025728  \\ \hline
	URYDR & 0.13055209877376 & -0.00955259258304  \\ \hline
	URYVL & 0.12149950412928 & 0.00221355374016  \\ \hline
	URYVR & 0.12099950619072 & 0.00710320988928  \\ \hline
	VA1 & -0.0026943210048 & 0.0247387654272  \\ \hline
	VA10 & -0.8088615384768 & 0.03357538460928  \\ \hline
	VA11 & -0.7857784617216 & 0.02785230769536  \\ \hline
	VA12 & -0.8911784615232 & 0.00842564102208  \\ \hline
	VA2 & -0.07020307689408 & 0.0366276922848  \\ \hline
	VA3 & -0.14498461539072 & 0.03815384615232  \\ \hline
	VA4 & -0.23960615386752 & 0.03968000001984  \\ \hline
\end{tabular}
\end{table}
\end{center}
	
\begin{center}
\begin{table}[h]
\centering
\begin{tabular}{ | l | l | l |}
\hline
	\textbf{Neuron Name} & \textbf{X} & \textbf{Y} \\ \hline
	VA5 & -0.34338461537088 & 0.04273230769536  \\ \hline
	VA6 & -0.44563692306624 & 0.04273230769536  \\ \hline
	VA7 & -0.54636307689408 & 0.04273230769536  \\ \hline
	VA8 & -0.6257230767552 & 0.03968000001984  \\ \hline
	VA9 & -0.71424 & 0.0366276922848  \\ \hline
	VB1 & 0.01494123459264 & 0.02400395061312  \\ \hline
	VB10 & -0.7386584615232 & 0.0366276922848  \\ \hline
	\rowcolor{lightgray} VB11 & -0.8180184613248 & 0.03357538460928  \\ \hline
	VB2 & 0.0293925925632 & 0.02351407405056  \\ \hline
	\rowcolor{lightgray} VB3 & -0.08088615384768 & 0.0366276922848  \\ \hline
	\rowcolor{lightgray} VB4 & -0.16024615382784 & 0.03815384615232  \\ \hline
	\rowcolor{lightgray} VB5 & -0.25181538462912 & 0.04120615382784  \\ \hline
	\rowcolor{lightgray} VB6 & -0.3693292307616 & 0.04273230769536  \\ \hline
	\rowcolor{lightgray} VB7 & -0.45632000001984 & 0.04273230769536  \\ \hline
	\rowcolor{lightgray} VB8 & -0.56315076925824 & 0.04120615382784  \\ \hline
	\rowcolor{lightgray} VB9 & -0.6364061537088 & 0.04120615382784  \\ \hline
	VC1 & -0.1663507692384 & 0.03815384615232  \\ \hline
	\rowcolor{lightgray} VC2 & -0.2609723077152 & 0.04120615382784  \\ \hline
	VC3 & -0.39680000001984 & 0.04273230769536  \\ \hline
	VC4 & -0.49752615384768 & 0.03815384615232  \\ \hline
	VC5 & -0.53262769232448 & 0.04273230769536  \\ \hline
	VC6 & -0.6531938458944 & 0.04120615382784  \\ \hline
	VD1 & -0.02179950617088 & 0.02424888886464  \\ \hline
	\rowcolor{lightgray} VD10 & -0.77376 & 0.0366276922848  \\ \hline
	\rowcolor{lightgray} VD11 & -0.8485415386752 & 0.03357538460928  \\ \hline
	\rowcolor{lightgray} VD12 & -0.8684451284544 & 0.01751897436864  \\ \hline
	\rowcolor{lightgray} VD13 & -0.9147384613248 & 0.0001589743488  \\ \hline
	VD2 & -0.03086222225088 & 0.02424888886464  \\ \hline
	\rowcolor{lightgray} VD3 & -0.13430153843712 & 0.03815384615232  \\ \hline
	\rowcolor{lightgray} VD4 & -0.21976615382784 & 0.04120615382784  \\ \hline
	\rowcolor{lightgray} VD5 & -0.31438769230464 & 0.04273230769536  \\ \hline
	\rowcolor{lightgray} VD6 & -0.40595692310592 & 0.04425846156288  \\ \hline
	\rowcolor{lightgray} VD7 & -0.4883692307616 & 0.04120615382784  \\ \hline
	\rowcolor{lightgray} VD8 & -0.5967261541056 & 0.04273230769536  \\ \hline
	\rowcolor{lightgray} VD9 & -0.6882953843712 & 0.03815384615232  \\ \hline
\end{tabular}
\end{table}
\end{center}

\end{document}